\documentclass[10pt,conference]{IEEEtran}

\usepackage{graphicx}
\usepackage{amsfonts,amssymb,amsmath,alltt,amsthm}
\usepackage{url}
\usepackage{color}
\newenvironment{ttbox}{\begin{alltt}\ttbraces\small\tt}%
                      {\end{alltt}}
\def\ttbraces{\let\.=\nobreak\chardef\{=`\{\chardef\}=`\}\chardef\|=`\\}

\newcommand{\red}[1]{{\textcolor{red}{#1}}}

\newcommand\ttand{\mbox{{$\land$}}}
\newcommand\ttor{\mbox{{$\lor$}}}
\newcommand\ttfun{\mbox{{$\Rightarrow$}}}

\newcommand\ttcup{\mbox{{$\cup$}}}

\newcommand\ttimp{\mbox{{$\longrightarrow$}}}
\newcommand\ttequiv{\mbox{{$\equiv$}}}
\newcommand\ttexists{\mbox{{$\exists$}}}
\newcommand\ttforall{\mbox{{$\forall$}}}
\newcommand\ttneg{\mbox{{$\neg$}}}
\newcommand\ttneq{\mbox{{$\neq$}}}
\newcommand\ttin{\mbox{{$\in$}}}
\newcommand\ttnin{\mbox{{$\notin$}}}
\newcommand\ttImp{\mbox{{$\Longrightarrow$}}}
\newcommand\ttlam{\mbox{\( \lambda \)}}
\newcommand\tttimes{\mbox{\( \times \)}}
\newcommand\ttsigma{\mbox{{$\sigma$}}}

\newcommand{\ttcalN}[1]{\mbox{{${\mathcal{N}}_{\texttt{#1}}$}}} 
\newcommand\ttattand[1]{\mbox{{$\oplus_{\wedge}^{#1}$}}}
\newcommand\ttattor[1]{\mbox{{$\oplus_{\vee}^{#1}$}}}
\newcommand\ttatI{\mbox{\( @_G \)}}

\newcommand\ttrelI{\mbox{{$\to$}}}
\newcommand\ttrelIstar{\mbox{{$\to^*$}}}
\newcommand\ttrel[1]{\mbox{{$\to_{#1}$}}}
\newcommand\ttrelstar[1]{\mbox{{$\to_{#1}^*$}}}

\newcommand\ttsubseteq{\mbox{{$\subseteq$}}}
\newcommand\ttsupseteq{\mbox{{$\supseteq$}}}

\newcommand\ttvdash{\mbox{{$\vdash$}}}
\newcommand\ttupdownarrow{\mbox{{$\Updownarrow$}}}
\newcommand\ttmref[1]{\mbox{{$\sqsubseteq_{#1}$}}}
\newcommand\ttmeref{\ttmref{\mathcal{E}}}

\newcommand\ttecal{\mbox{$\mathcal{E}$}}
\newcommand\ttimg{\mbox{$\triangleleft$}}

\begin{document}

%% Title information
\title{A Formal Development Cycle for Security Engineering in Isabelle}         %% [Short Title] is optional;
                                        %% when present, will be used in
                                        %% header instead of Full Title.
%\titlenote{with title note}             %% \titlenote is optional;
                                        %% can be repeated if necessary;
                                        %% contents suppressed with 'anonymous'
%\subtitle{Subtitle}                     %% \subtitle is optional
%\subtitlenote{with subtitle note}       %% \subtitlenote is optional;
                                        %% can be repeated if necessary;
                                        %% contents suppressed with 'anonymous'

%% Author information
%% Contents and number of authors suppressed with 'anonymous'.
%% Each author should be introduced by \author, followed by
%% \authornote (optional), \orcid (optional), \affiliation, and
%% \email.
%% An author may have multiple affiliations and/or emails; repeat the
%% appropriate command.
%% Many elements are not rendered, but should be provided for metadata
%% extraction tools.

\author{\IEEEauthorblockN{Florian Kamm\"uller}
\IEEEauthorblockA{Middlesex University London, UK\\
f.kammueller@mdx.ac.uk}}

%% Author with single affiliation.
%\author{Florian Kamm\"uller}
%\authornote{}          %% \authornote is optional;
                                        %% can be repeated if necessary
%\orcid{0000-0001-5839-5488}             %% \orcid is optional
%\affiliation{
%  \position{Position1}
%  \department{Department1}              %% \department is recommended
%  \institution{Middlesex University London and TU Berlin}            %% \institution is required
%  \streetaddress{Street1 Address}
%  \city{London and Berlin}
%  \state{State1}
%  \postcode{Post-Code1}
%  \country{UK and Germany}                    %% \country is recommended
%}
%\email{f.kammueller@mdx.ac.uk}          %% \email is recommended
%\acmConference[arxive]{ }{2020}{ }
%\authornotemark{ }
%% Abstract
%% Note: \begin{abstract}...\end{abstract} environment must come
%% before \maketitle command
%% \maketitle
%% Note: \maketitle command must come after title commands, author
%% commands, abstract environment, Computing Classification System
%% environment and commands, and keywords command.
\maketitle

\begin{abstract}
In this paper, we show a security engineering process based on
a formal notion of refinement fully formalized in the proof assistant 
Isabelle.
This Refinement-Risk Cycle focuses on attack analysis and 
security refinement supported by interactive theorem proving.
Since we use a fully formalized model of infrastructures with actors and 
policies we can support a novel way of formal security refinement for system 
specifications. This formal process is built practically as an 
extension to the Isabelle Infrastructure framework with attack trees. 
We define a formal notion of refinement on infrastructure models. 
Thanks to the formal foundation of Kripke structures and branching time 
temporal logic in the Isabelle Infrastructure framework, these stepwise 
transformations can be interleaved with attack tree analysis thus 
providing a fully formal security engineering framework.
The process is illustrated on an IoT healthcare case study introducing
GDPR requirements and blockchain.
\end{abstract}

%% 2012 ACM Computing Classification System (CSS) concepts
%% Generate at 'http://dl.acm.org/ccs/ccs.cfm'.
%\begin{CCSXML}
%<ccs2012>
%<concept>
%<concept_id>10011007.10011006.10011008</concept_id>
%<concept_desc>Software and its engineering~General programming languages</concept_desc>
%<concept_significance>500</concept_significance>
%</concept>
%<concept>
%<concept_id>10003456.10003457.10003521.10003525</concept_id>
%<concept_desc>Social and professional topics~History of programming languages</concept_desc>
%<concept_significance>300</concept_significance>
%</concept>
%</ccs2012>
%\end{CCSXML}

%\ccsdesc[500]{Software and its engineering~General programming languages}
%\ccsdesc[300]{Social and professional topics~History of programming languages}
%% End of generated code

%% Keywords
%% comma separated list
%\keywords{Isabelle theorem proving, Security and Privacy, IoT}  %% \keywords are mandatory in final camera-ready submission

\section{Introduction}
Security is a notoriously difficult property for system development because 
it is not compositional: given secure components, a system created from those 
components is not necessarily secure. Therefore, the usual divide-and-conquer 
approach from system and software design does not apply for security 
engineering. 
At the same time, it is mandatory for the design of secure systems to introduce
security in the early phases of the development since it cannot be easily 
``plugged in'' at later stages. However, even if security is introduced in 
early phases, a classical stepwise development of refining abstract 
specifications by concretizing the design does not preserve security 
properties. Take for example, the implementation of sending a message from 
a client A to a server B such that the communication is encrypted to protect 
its content. In the abstract system specification we do not consider a concrete
protocol nor the architecture of the client and server. Using common refinement
methods from software engineering provides a possible implementation by passing
the message from a client system AS connected by a secure
channel to a server system BS. However, this implementation does not exclude 
that other processes running on either AS or BS can eavesdrop on the cleartext 
message because the confidentiality protection is only on the secure channel 
from AS to BS. This example is a simple illustration of what is known as the 
security refinement paradox \cite{jac:89}.
Why is security so hard? A simple explanation is that it talks about negative 
properties: something (loss or damage of information or functionality) must 
not happen. Negation is also in logic a difficult problem as it needs 
exclusion of possibilities. If the space to consider is large, this proof 
can be hard or infeasible. In security, the attacks often come from 
``outside the model''. That is, for a given specification we can prove 
some security property and yet an attack occurs which uses a fact or 
observation or loophole that just has not been considered in the model. 
This known practical attack problem is similar to the refinement paradox. 
Intuitively, the attacker exploits a refinement of the system that has not 
been taken into account in the specification but is actually part of the 
real system (an implementation of the specification). In the above example, 
the real system allows that other applications can be run on the client within 
the security boundary. This additional feature of multi-processing systems 
has not been taken into account in the abstract specification in the above example 
where we considered processes and systems -- the client and the server -- as abstract 
entities without distinguishing the features of their internal architectures.
%However, these intricacies of security refinement and preservation of 
%security do not occur so much, if we lift the negation. Instead of considering 
%the negative security property, we double negate and consider ``not secure''. 
%In other words, we contemplate attack possibilities. The working hypothesis of 
%this paper is that we should have stronger refinement properties when 
%considering attacks. What we want to examine is whether attacks can be 
%refined for given system models. We will do this by first providing a 
%formal model of attacks using the well known attack tree concept but underlaying it with a formal foundation by providing a logical proof theory for it. Refinement is in fact the natural way of proving an attack on a given system by applying the calculus to find a concrete way that realizes an abstract attack. The attack refinement shows how the concrete implementation must look like in order to enable the attack. Thus, ultimately, we find a negative -- not secure 
%implementation of an abstract system -- by letting the
%attack lead the way of the refinement.
%Beyond this, there is a range of others important observations concerning 
%the relation between attack trees, infrastructure representations, and 
%modelchecking in Isabelle that we present in this paper and whose conception, 
%construction, and demonstration represents our contribution.

Sadly, 100\% security is not achievable, therefore, the next best thing is to
find ways to gradually improve the security of a system. This should ideally be done
at design time since a specification can be changed while changing a system is expensive.
A pragmatic way of engineering secure systems is to pursue the identification of security
goals as part of a security requirements engineering process while complementing
this with the establishment of an attacker model.
To this end, we propose a formal process of system development that focuses
on attack analysis and security refinement in one integrated interactive process
interleaving system development with attack analysis using attack trees \cite{Schneier.102}. 
To support this ambitious goal, the current paper pulls together strands of previous
works, for the first time succeeding in combining attack analysis and system refinement
in one consistent automated framework illustrating it on a complex case study.
% fully integrating powerful logical specification 
%of high-level system designs and policices
%with machine assisted proof of security properties.
The paper presents a generic theory of refinement in Isabelle that manifests 
linking the notion of attack trees with state based system refinement based on a Kripke 
semantics and temporal logic. The security refinement is illustrated on an IoT healthcare
example that is entirely formalized in Isabelle on top of the underlying theory.

%The means to enable us to integrate security at such an early phase at design time is that
%we use powerful logical specification of high-level system designs and machine assisted 
%proof of security properties to enhance the process. 
%This Refinement-Risk Loop interleaves attack tree analysis with model 
%transformation and is fully formalised within the interative Isabelle framework. 

The contributions of this paper are:
(a) we present a fully formalized process of security refinement for infrastructure
systems using a notion of refinement  and derive useful theory for it.
The resulting Refinement-Risk Cycle
integrates formal system development with risk analysis by attack trees;
%an iterative process of system specifications 
%with attack tree analysis that incrementally refines a system specification;
(b) we illustrate the process by showing the development of an IoT healthcare 
system from the CHIST-ERA project SUCCESS \cite{suc:16} exhibiting security attacks
and formally refining the system specification step-by-step.
An earlier workshop paper \cite{kam:19a}, already introduced the Refinement-Risk
Cycle but only informally exhibiting the IoT healthcare example. The current paper subsumes this workshop paper by defining a formal process of refinement.
It thus formalizes a process of security engineering. 
Interestingly, although the system itself had also been formalised in the informal 
precursor \cite{kam:19a}, the earlier specification contains subtle design errors that make 
a secure refinement impossible.
These errors have now been identified by applying the formal refinement that 
constitutes the fundamental core of the Refinement-Risk cycle.
Thus, the current addition of the foundation of the Refinement-Risk cycle is a 
contribution that not only largely extends \cite{kam:19a} but by scrutinizing the 
example establishes the validity of the security refinement and proves 
its valor. As a further additional contribution, we finally present the analysis of the
error correction.

In the remainder of this section we briefly summarize the underlying 
Isabelle Infrastructure framework including Kripke structures, CTL and Attack 
Trees that we used as the foundation for the current work. A detailed
account is contained in the Appendix.
Next, we present the Refinement-Risk-Cycle (RR-Cycle) and in particular the
formal notion of refinement that allows security refinement 
(Section \ref{sec:rrloop}).
%and an overview of 
We then illustrate this process on the application to the case study 
first giving an overview and initial model (Section \ref{sec:appini})
before providing technical details of the cycle's application by the stepwise 
system refinement steps triggered by attacks 
(Section \ref{sec:apploop}). We finally present the design errors that could be 
identified (Section \ref{sec:errors}) before we conclude in Section \ref{sec:concl}.
All developments and the application to the case study are formalised 
in Isabelle. %(see accompanying sources).
The sources are available online \cite{kam:19git}.

\subsection{Kripke structures, CTL, and Attack Trees}
\label{sec:intro}
\begin{figure}[h!]
\begin{center}
\includegraphics[scale=.4]{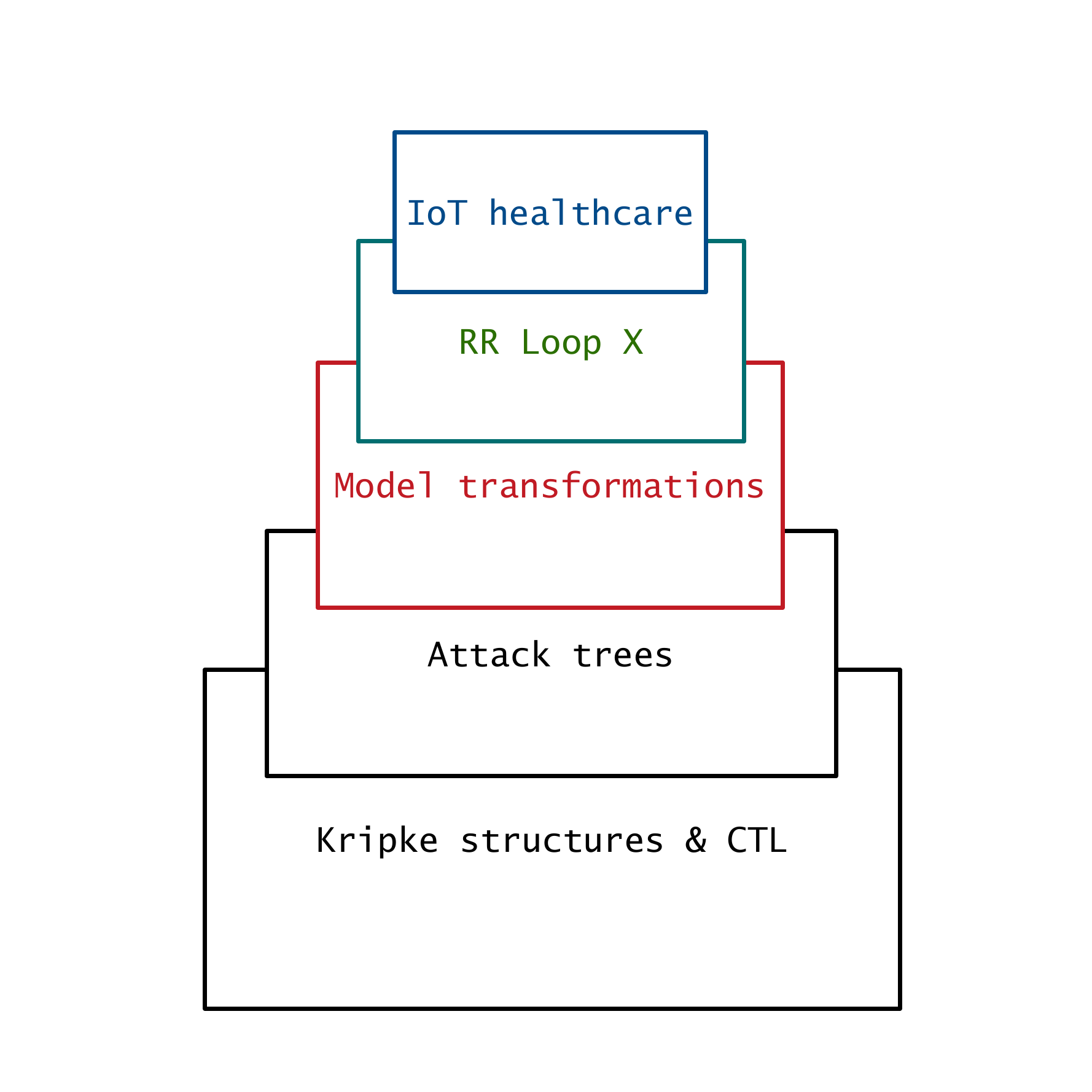}
\end{center}
%\vspace{-.5cm}
\caption{Generic framework for infrastructures with refinement.}
\label{fig:theorystruc}
\end{figure}

Figure \ref{fig:theorystruc} gives an overview of the Isabelle Infrastructure 
framework with its layers of object-logics -- each level below embeds the one
above showing the novel contribution of this paper in colours on the top.

In the course of various extensions (detailed in the Appendix), 
the Isabelle framework has been restructured such that it
is now a general framework for the state-based security analysis of infrastructures with 
policies and actors. Temporal logic and Kripke structures build the foundation. 
Meta-theoretical results have been established to show equivalence between attack trees 
and CTL statements \cite{kam:18b}. 
%As part of the meta-theory Correctness and Completeness have been 
%proved in Isabelle \cite{kam:18b} and can be used to navigate between CTL and attack 
%trees to find attacks. 
This foundation provides a generic notion of state transition on which attack trees and
temporal logic can be used to express properties. 
The main notions used in this paper are:
\begin{itemize}
\item {Kripke structures and state transitions:}\\ 
Using a generic state transition relation $\mapsto$, Kripke structures
are defined as a set of states reachable by $\mapsto$ from an initial
state set, for example
\begin{ttbox}
Kripke \{t. \ttexists i \ttin I. i \ttrelIstar t\} I
\end{ttbox} 
\item {CTL statements:}\\ 
For example, we can write 
\begin{ttbox}
K \ttvdash {\sf EF} s
\end{ttbox}
to express that in Kripke structure \texttt{K} there is a path on which
the property \texttt{s} (a set of states) will eventually hold.
\item {Attack trees:} \\
The datatype of attack trees has three constructors: 
$\oplus_\vee$ creates or-trees and $\oplus_\wedge$ creates 
and-trees.
And-attack trees $l \ttattand s$ and or-attack trees $l \ttattor s$ 
consist of a list of sub-attacks -- again attack trees. The third constructor 
creates a base attack as a pair of state sets written \texttt{\ttcalN{(I,s)}}.
For example, a two step and-attack leading from state set \texttt{I} via
\texttt{si} to \texttt{s} is expressed as
\begin{ttbox}
\ttvdash [\ttcalN{(I,si)},\ttcalN{(si,s)}]\ttattand{\texttt{(I,s)}}
\end{ttbox}
%\begin{ttbox}
%\ttvdash [\ttcalN{(I,s)},\ttcalN{(s,s')}]\ttattand{(I,s')}
%\end{ttbox}
\item {Attack tree refinement, validity and adequacy:}\\
Attack trees have their own refinement (not to be mixed up with the
model transformation presented in this paper). An abstract attack tree
may be refined by spelling out the attack steps until a valid attack
is reached: \texttt{\ttvdash A :: (\ttsigma :: state) attree)}.
The validity is defined constructively (code is generated from it)
and its adequacy with respect to a formal semantics in CTL is proved
and can be used to facilitate actual application verification as demonstrated 
her in the stepwise system refinements.
\end{itemize}

In this paper, we present an extension of this formal process introducing
refinement of Kripke structures. It refines a system model based on a 
formal definition of a combination of trace refinement and structural 
refinement. The definition allows to prove property preservation results 
crucial for an iterative development process.
%an integrated process
%process showing how 
The refinements of the system specification  can be interleaved with attack 
analysis while security properties can be proved in Isabelle. In each iteration
security qualities are accumulated while continuously attack trees scrutinize
the design.

\section{The Refinement-Risk-Cycle for Secure IoT System}
\label{sec:rrloop}
We first introduce the iterative process of refinement and attack tree analysis 
(the ``Refinement-Risk-Cycle'') providing an overview followed by the formal definition
in Isabelle and the resulting property preservation.

\subsection{Overview of RR-Cycle}
\begin{figure}
\begin{center}
\includegraphics[scale=.21]{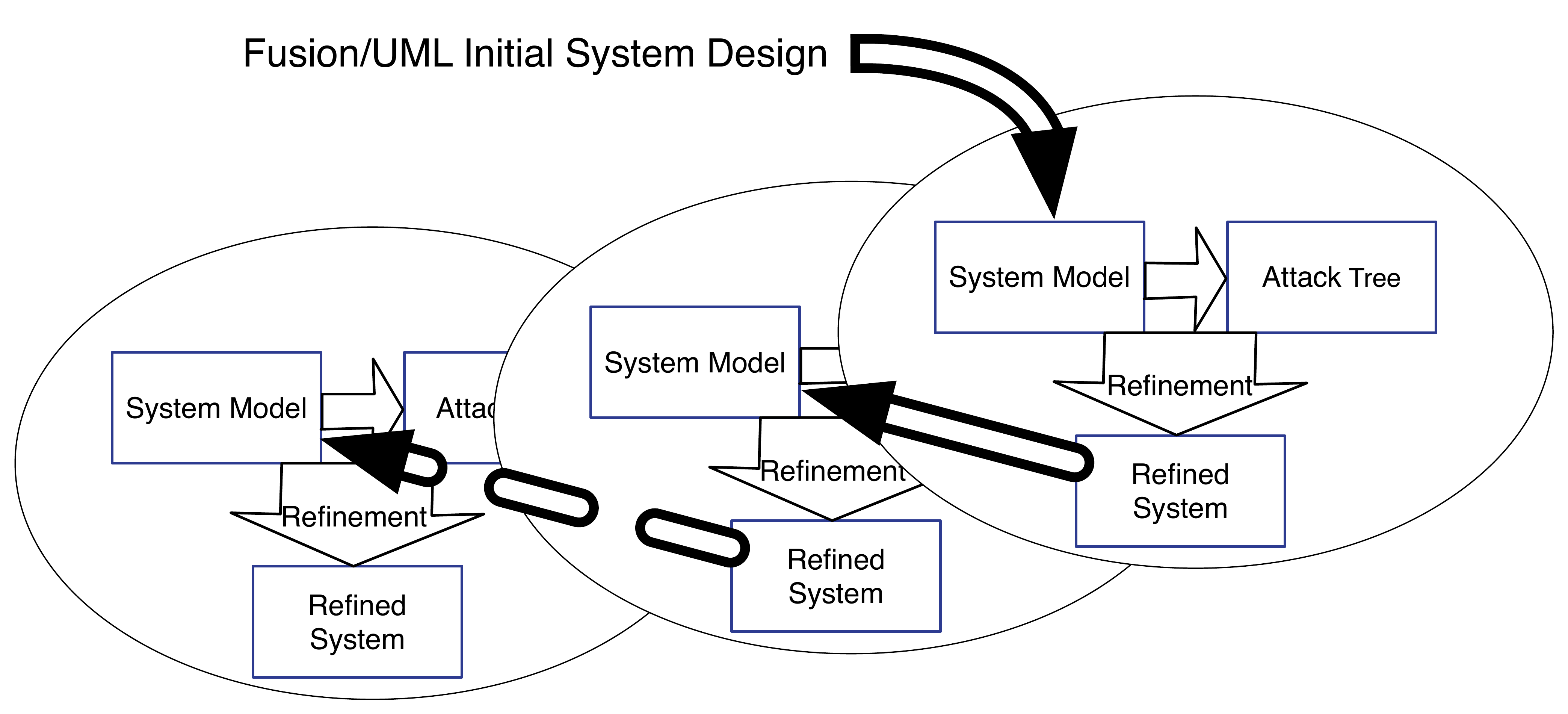}
\end{center}
\caption{Refinement-Risk-Cycle iterates design, risk analysis, and refinement}\label{fig:process}
%\vspace{-1cm}
\end{figure}

As an initial step, the Fusion/UML method serves to develop a system 
architecture from early requirements. 
This system architecture is translated into the Isabelle Infrastructure framework: 
actors in UML become Isabelle Infrastructure actors, UML system classes are represented 
by locations in the infrastructure graph, and the class attributes and pre- and postconditions 
of methods are formalised in the local and global policies.
The identification of attacks, using for example invalidation \cite{kp:14},
can then reveal paths of state transitions through the system model where the
global security policy is violated. In an iteration, these attack paths provide
details useful for refining the system specification by adding security controls,
for example, access control, privacy preservation, or blockchain. 
The addition of detail, however, may in turn introduce new vulnerabilities that lead to
new iterations of the process.
Security properties may be proved at each level of the iteration. They are
true for this abstraction level of the system model and remain true in the 
refined system. However, %as is known in the 
%``security paradox'': attacks mostly come from outside the model. Attacks 
new attacks may be found despite proved security. 
%If these attacks undermine the proven properties,
%it is because they use information not present in the model.
%But this yields the key to finding a refinement: introducing a level of detail 
%that enables a formal or computational representation of the details used in the attack 
%incarnates the next refinement.
The Refinement-Risk Cycle process is graphically depicted in Figure \ref{fig:process}.
%\vspace{.1cm}
%We now provide a formal definition of the core of the Refinement in
%the RR-Loop by introducing the concept of model transformations
%formally as generic layer on top of Kripke structures.
%This extension of our Infrastructure framework %will then enable to
%enables capturing model transformation formally for any type of 
%Kripke structure state and state transition relation, the concrete infrastructures
%we define for IoT being just an example. 
%This is due to the fact that we have designed and constructed the 
%Isabelle Infrastructure framework generically.
%Moreover, the model transformation supports property preserving system refinement.

\subsection{Refinement}
\label{sec:transform}
Intuitively, a refinement changes some aspect of the type of
the state, for example, replaces a data type by a richer datatype or
restricts the behaviour of the actors. The former is expressed directly 
by a mapping of datatypes, the latter is incorporated into the state
transition relation of the Kripke structure that corresponds to the 
transformed model.
%This relationship between the Kripke structure can be better understood
%visually (see Figure \ref{fig:modtrans})
In other words, we can encode a refinement within our framework
as a relation on Kripke structures that is parametrized additionally by
a function that maps the refined type to the abstract type.
The direction ``from refined to abstract'' of this type mapping may seem 
curiously counter-intuitive. However, the actual refinement is given by the 
refinement that uses this function as an input. The refinement
then refines an abstract to a more concrete system specification. 
The additional layer for the refinement 
%(the red box in Figure \ref{fig:theorystruc})
can be formalised in Isabelle as a second\footnote{The first refinement 
relation in this framework is on attack trees summarized in Section \ref{sec:at}.} 
refinement relation 
$\sqsubseteq_{\mathcal{E}}$. 
The relation \texttt{mod\_trans} is typed as a relation over triples --
a function from a threefold Cartesian product to \texttt{bool}, the 
type containing true and false only.  
The type variables $\sigma$ and $\sigma'$ input to the type constructor 
\texttt{Kripke} represent the abstract state type and the concrete state type. 
Consequently, the middle element of the triples selected by the relation 
\texttt{mod\_trans} is a function of type $\sigma' \Rightarrow \sigma$ 
mapping elements of the refined state to the abstract state.
The expression in quotation marks after the type is again the
infix syntax in Isabelle that allows the definition of mathematical notation
instead of writing \texttt{mod\_trans} in prefix manner. This nicer infix
syntax is already used in the actual definition.
Finally, the arrow \texttt{\ttImp} is the implication of Isabelle's  meta-logic
while $\ttimp$ is the one of the {\it object} logic HOL. 
They are logically equivalent but of different
types: within a HOL formula $P$, for example, as below $\forall x. P \ttimp Q$, only the implication $\ttimp$
can be used.
\begin{ttbox}
 mod_trans ::  (\ttsigma Kripke \tttimes (\ttsigma' \ttfun \ttsigma) \tttimes \ttsigma' Kripke) 
               \ttfun bool                  ("_ \ttmref{(\_)} _")
  K \ttmeref K' \ttequiv \ttforall s' \ttin states K'. \ttforall s \ttin init K'. 
             s \ttrelstar{\sigma'} s' \ttimp \ttecal(s) \ttin init K 
             \ttand \ttecal(s) \ttrelstar{\sigma} \ttecal(s')
\end{ttbox}
The definition of \texttt{K \ttmeref\, K'} states that for any state $s$ 
of the refined Kripke structure that can be reached by the state transition
in zero or more steps from an initial state $s_0$ of the refined Kripke 
structure, the mapping ${\mathcal E}$ from the refined to the abstract 
model's state must preserve this reachability, i.e., the image of
$s_0$ must also be an initial state and from there the image of $s$
under ${\mathcal E}$ must be reached with $0$ or $n$ steps.

\subsection{Property Preserving System Refinement}
A first direct consequence of this definition is the following lemma
where the operator \texttt{$\ttimg$} in \texttt{\ttecal\ttimg(init K')}
represents function image, that is the set, $\{\ttecal(x). x \in \texttt{init K'}\} $.
\begin{ttbox}
{\bf{lemma}} init_ref: K \ttmeref K' \ttImp \ttecal\ttimg(init K') \ttsubseteq init K
\end{ttbox}
A more prominent consequence of the definition of refinement 
is that of property preservation. Here, we show that refinement preserves the
CTL property of ${\sf EF} s$ which means that a reachability property true in the
refined  model \texttt{K'} %has also been 
is already true in the abstract model.
A state set $s'$ represents a property %since we use 
in the predicate transformer view of properties as sets of states. 
The additional condition on initial states ensures that we cannot ``forget'' them. 
%initial states.
%The theorem states that, if the 
%property \texttt{{\sf EF} s'} is true in the concrete model its  has been true 
%can be reached from the initial state in 
%\texttt{K'} then it is also possible  to reach the image of this property
% in the abstract Kripke structure \texttt{K}.
\begin{ttbox}
{\bf{theorem}} prop_pres: 
   K \ttmeref K'  \ttImp init K \ttsubseteq \ttecal\ttimg(init K') \ttImp
   \ttforall s' \ttin Pow(states K'). K' \ttvdash {\sf EF} s' 
              \ttimp K \ttvdash {\sf EF} (\ttecal\ttimg(s'))
\end{ttbox}
It is remarkable, that our definition of refinement by Kripke 
structure refinement entails property preservation and makes it possible 
to prove this as a theorem in Isabelle once for all, i.e., as a meta-theorem.
However, this is due to the fact that our generic definition of state transition
allows to explicitly formalise such sophisticated concepts like reachability.
For practical purposes, however, the proof obligation of showing that
a specific refinement is in fact a refinement is rather complex
justly because of the explicit use of the transitive closure of the state
transition relation.
In most cases, the refinement will be simpler. Therefore, we offer
additional help by the following theorem that uses a stronger characterisation
of Kripke structure refinement and shows that our refinement follows
from this.
\begin{ttbox}
{\bf{theorem}} strong_mt: 
\ttecal\ttimg(init K') \ttsubseteq init K \ttand s \ttrel{\sigma'} s' \ttimp \ttecal(s) \ttrel{\sigma} \ttecal(s') 
\ttImp K \ttmeref K'
\end{ttbox}
This simpler characterisation is in fact a stronger one: we could have $s \ttrel{\sigma'} s'$ 
in the refined Kripke structure \texttt{K'} and $\neg(\ttecal(s) \ttrel{\sigma} \ttecal(s'))$
but neither $s$ nor $s'$ are reachable from initial states in \texttt{K'}.
For cases, where we want to have the simpler one-step proviso but still need 
reachability we provide a slightly weaker version of \texttt{strong\_mt}.
\begin{ttbox}
{\bf{theorem}} strong_mt':  
\ttecal\ttimg(init K') \ttsubseteq init K \ttand (\ttexists s0 \ttin init K'. s0  \ttrelIstar s)
 \ttand s \ttrel{\sigma'} s' \ttimp \ttecal(s) \ttrel{\sigma} \ttecal(s') \ttImp K \ttmeref K'
\end{ttbox}

This idea of property preservation coincides with the classical idea of
trace refinement as it is given in process algebras like CSP. In this view,
the properties of a system are given by the set of its traces. Now, a refinement
of the system is given by another system that has a subset of the traces of the 
former one.
Although the principal idea is similar, we greatly extend it since our notion
additionally incorporates refinement. Since we include a state map 
\texttt{\ttsigma'\ttfun \ttsigma} in our refinement map, we additionally
allow structural refinement: the state map generalises the basic idea of
trace refinement by traces corresponding to each other but allows additionally
an exchange of data types. 
As we see in the application to the case study, the refinement steps may
sometimes just specialise the traces: in this case the state map 
\texttt{\ttsigma'\ttfun \ttsigma} is just identity.

\section{Applying RR-Cycle to IoT Healthcare Example}
\label{sec:appini}
We now first give a tabular overview of the steps taken for the case study. 
Following the RR-Cycle, we have modelled and analysed the IoT healthcare application
in four iterations summarised in %the following table.
the table in Figure \ref{tab:rrlapp}.

How each of these models is refined in each iteration, 
as well as the attack trees that exhibit vulnerabilities, is discussed in the 
following sections as indicated in the last column of the table in Figure \ref{tab:rrlapp}.
%The technical details of these steps are discussed next.

\begin{figure}
\noindent%
\begin{tabular}{|p{3.2cm}|p{2.8cm}|p{1.3cm}|}
\cline{1-3}
{\bf System} & {\bf Attack} & {\bf Where }\\
\hline\hline
Initial Fusion system home-cloud-hospital & Eve can perform action get at cloud & 
   Sections \ref{sec:appini}--\ref{sec:getatt}\\
\hline 
\multicolumn{3}{|c|}{\it Refinement-Risk-Cycle Iteration 1} \\ 
\hline
Access control by DLM labels & Eve can perform action eval at cloud; changes label to her own & 
   Sections \ref{sec:label}, \ref{sec:atteval} \\
\hline 
\multicolumn{3}{|c|}{\it Refinement-Risk-Cycle Iteration 2} \\
\hline
Privacy preserving functions type \texttt{label\_fun} & Eve puts Bob's data labelled as her own & 
Sections \ref{sec:rrtwo}, \ref{sec:attput}\\ 
\hline 
\multicolumn{3}{|c|}{\it Refinement-Risk-Cycle Iteration 3} \\
\hline
Global blockchain & Eve is an insider impersonating the blockchain controller & 
Sections \ref{sec:ppfun}--\ref{sec:guar}\\
\hline 
\multicolumn{3}{|c|}{\it Refinement-Risk-Cycle Iteration 4} \\
\hline
Consensus (for example Nakamoto) blockchain &  no attack known yet & Section \ref{sec:rrfour}\\
\hline 
\end{tabular}
\caption{Iterated application of Refinement-Risk(RR)-Cycle}\label{tab:rrlapp}
\end{figure}

\subsection{Initial Step: Fusion/UML for System Architecture}
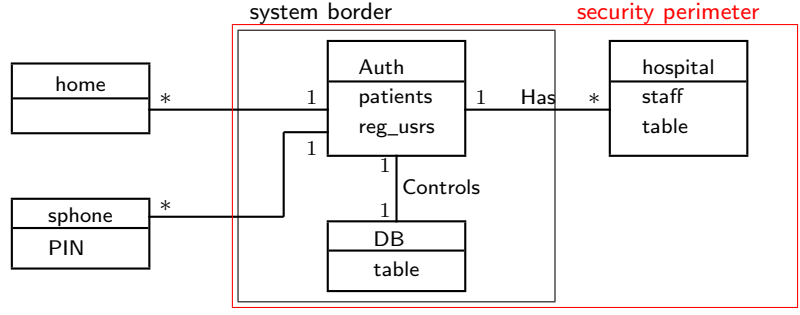
\begin{figure*}
\begin{center}
\setlength{\unitlength}{0.15mm}
{\small
\begin{picture}(750,200)
% figure
\thicklines
\put(0,120){\framebox(120,60){\footnotesize $\begin{array}{l}{{\sf home}}\\
                                                 \    \end{array}$}
}
\put(0,150){\line(1,0){120}}
% link
%\path(120,120)(240,120)(240,100)(280,100)
\put(120,140){\line(1,0){160}}
%\put(240,140){\line(0,-1){20}}
%\put(240,120){\line(1,0){40}}
\put(130,145){$*$}
\put(260,145){\footnotesize$1$}

\thicklines
\put(0,0){\framebox(120,60){\footnotesize $\begin{array}{l}{\small {\sf sphone}}\\[1ex]
                                                 {\sf PIN} \end{array}$}
}
\put(0,35){\line(1,0){120}}
% link
%\path(120,40)(240,40)(240,60)(280,60)
\put(120,45){\line(1,0){120}}
\put(240,45){\line(0,1){75}}
\put(240,120){\line(1,0){40}}
\put(130,50){$*$}
\put(260,100){\footnotesize$1$}

% system border
\thinlines
\put(200,-30){\framebox(280,240)}
\put(210,220){{\rm ${\sf system\ border}$}}
% security perimeter
\red{\put(195,-35){\framebox(500,250)}}
\put(500,220){{\rm ${\sf \red{security\ perimeter}}$}}

% controller
\thicklines
\put(280,100){\framebox(120,100){\footnotesize $\begin{array}{l}{\sf Auth}\\[.6ex]
                                                 {\sf patients} \\[.6ex]
                                                 {\sf reg\_usrs}\end{array}$}}

\put(280,165){\line(1,0){120}}
% links
\put(400,140){\line(1,0){130}}
\put(410,145){\footnotesize$1$}
\put(450,145){\footnotesize ${\sf Has}$}
\put(510,145){$*$}
%\put(650,140){\line(1,0){130}}
% collaborators
%database
\thicklines
\put(280,-20){\framebox(120,60){\footnotesize $\begin{array}{l}{\sf DB}\\[.6ex]
                                                {\sf table} \end{array}$}}
\put(280,15){\line(1,0){120}}
\put(340,40){\line(0,1){60}}
\put(325,45){\footnotesize$1$}
\put(325,85){\footnotesize$1$}
\put(345,65){\footnotesize${\sf Controls}$}
% hospital
\put(530,100){\framebox(120,100){\footnotesize $\begin{array}{l}{\sf hospital}\\[.6ex]
                                                 {\sf staff}\\[.6ex]
                                                 {\sf table}\end{array}$}}
\put(530,165){\line(1,0){120}}

\end{picture}
}
\end{center}
\vspace{.5cm}
\caption[]{System class model for IoT healthcare system %\TODO{(some associations and 
%hospital on the right inside system border still missing)}
}\label{fig:arch}
%\hrule
%\vspace{-.5cm}
\end{figure*}

The Fusion/UML process for object oriented design and analysis has been used to derive
a system design for the application scenario.
%cut short the full 
For reasons of conciseness, we omit here the details %application of the Fusion/UML process
presenting just one of the main outcomes of the analysis process: the system class 
model as depicted in Figure \ref{fig:arch}.
Note that, within the security perimeter, only the cloud server and the 
connected hospital (or other client institutions) are situated. The smartphone and 
the home server feature as data upload devices and the smartphone additionally as 
a control device that is included in some of the use cases.
This is a consequence of the GDPR \cite{kam:18a} requirements which are thus immediately
settled in the initial architecture.
% Why?
%\vspace{1cm}
Another result of the Fusion/UML analysis along with this system architecture 
is a set of operation schemas based on the system class model, additional use cases
and object collaborations. For details see \cite{kop:18}.
%These are omitted here but can be requested from the author.
%The 
%A major observation in the system architecture depicted in Figure \ref{fig:arch}
%is that the security perimeter stretches over two separate distributed systems.
%This is not common for Fusion/UML designs and thus no clear recepy for implementation.
%We propose in this paper the use of a distributed ledger to guarantee the security 
%and privacy goals across distributed system parts. To provide a general solution
%for this we offer the extension of the Isabelle Infrastructure framework by
%decentralized labels and a ledger. Thereby, a system design can be formally 
%specified to guarantee privacy in the sense of GDPR if it adheres to this
%specification.

\subsection{Infrastructures, Policies, and Actors}
\label{sec:infra}
% Adapted from CSF paper
The Isabelle Infrastructure framework supports the representation of infrastructures 
as graphs with actors and policies attached to nodes. These infrastructures 
are the {\it states} of the Kripke structure. % for the attack trees.

The transition between states is triggered by non-parametrized
actions \texttt{get}, \texttt{move}, \texttt{eval}, and \texttt{put} 
executed by actors. 
Actors are given by an abstract type \texttt{actor} and a function 
\texttt{Actor} that creates elements of that type from identities 
(of type \texttt{string} written \texttt{''s''} in Isabelle). 
Actors are contained in an infrastructure graph constructed by
\texttt{Lgraph} -- here the IoT healthcare case study example.
\begin{ttbox}
 ex_graph \ttequiv  Lgraph 
  \{(home,cloud), (sphone,cloud), (cloud,hospital)\}
    (\ttlam x. if x = home then \{''Patient''\} else 
      (if x = hospital then \{''Doctor''\} else \{\})) 
      ex_creds ex_locs
\end{ttbox}

%\begin{ttbox}
%{\bf datatype} igraph = 
%         Lgraph (location \tttimes location)set 
%                 location \ttfun identity set
%                 actor \ttfun (string set \tttimes string set)  
%                 location \ttfun string
%\end{ttbox}
%%                 ledger
This graph contains a set of location pairs representing the %structure 
topology of the infrastructure
as a graph of nodes and a function%\footnote{ We use the common $\lambda$-abstraction, 
%e.g. \texttt{\ttlam x. True}, to define functions with parameters, here the 
%function returning True for any input $x$.} 
that assigns a set of actor identities to each node (location) in the graph.
The last two graph components \texttt{ex\_creds} and \texttt{ex\_locs} 
are here abbreviated only (for the definitions see \cite{kam:18smc}).
%Also an infrastructure graph 
The function \texttt{ex\_creds} associates actors to a pair of string sets by
a pair-valued function whose first range component is a set describing the credentials 
in the possession of an actor and %the second component 
%is a set defining 
the roles the actor can take on; 
%Moreover, importantly in this context
%an infrastructure graph %assigns locations a string that 
\texttt{ex\_locs} defines the data residing at the component.
%and an element of type \texttt{acond}. This
%%type \texttt{acond} is defined as a set of labelled data representing a condition
%%on that data. 
%The final component of an \texttt{igraph} is the ledger.
Corresponding projection functions for each of the components of an 
infrastructure graph are provided; they are named \texttt{gra} for the actual 
set of pairs of locations, \texttt{agra} for the actor map, \texttt{cgra} for 
the credentials, and \texttt{lgra} for the %state of a location and the 
data at that location.
% and \texttt{ledgra} for the ledger.
%%and predicates \texttt{has} and 
%%\texttt{role} express that actors have credentials or that they 
%%can perform in specified roles while \texttt{isin} says that 
%%locations are in a specified state.

Infrastructures %are given by the following datatype that 
contain an infrastructure graph %of type \texttt{igraph} 
and a policy 
given by a function that assigns local policies over a graph to
all locations of the graph. 
\begin{ttbox}
 {\bf{datatype}} infrastructure = 
  Infrastructure  igraph 
                 [igraph, location] \ttfun policy set
\end{ttbox}
There are projection functions \texttt{graphI} and \texttt{delta} when applied
to an infrastructure return the graph and the policy, respectively.
For our healthcare example, the initial infrastructure contains
the above graph \texttt{ex\_graph} and the local policies defined shortly.
\begin{ttbox}
 hc_scenario \ttequiv Infrastructure 
                      ex_graph local_policies
\end{ttbox}
The function \texttt{local\_policies} gives the policy for each location \texttt{x} 
over an infrastructure graph \texttt{G} as a pair: the first element of this pair is 
a function specifying the actors \texttt{y} that are entitled to perform the actions 
specified in the set which is the second element of that pair.
%to illustrate the use of data labelling 
%and the ledger for consistent enforcement of GDPR data protection.
%%\footnote{The $\lambda$ in the definition is the usual lambda-operator of higher order logic that describes functions. 
%%For instance, the square function can be defined -- without giving it a name -- as $\ttlam x. x*x$.
%}.
\begin{ttbox}
 local_policies G x \ttequiv
 case x of 
   home \ttfun \{(\ttlam y. True, \{put,get,move,eval\})\}
 | sphone \ttfun 
  \{((\ttlam y. has G (y,''PIN'')), \{put,get,move,eval\})\} 
 | cloud \ttfun \{(\ttlam y. True, \{put,get,move,eval\})\}
 | hospital \ttfun 
     \{((\ttlam y. (\ttexists n. (n  \ttatI hospital) \ttand 
        Actor n = y \ttand has G (y, ''skey''))),
      \{put,get,move,eval\})\} 
 | _ \ttfun  \{\})
\end{ttbox}
Policies specify the expected behaviour of actors of an infrastructure. 
They are given by pairs of predicates (conditions) and sets of (enabled) actions.
%\begin{ttbox}
%{\bf type}_{\bf{synonym}} policy = ((actor \ttfun bool) \tttimes action set)
%\end{ttbox}
%The behaviour of actors is 
They are defined by the \texttt{enables} predicate:
an actor \texttt{h} is enabled to perform an action \texttt{a} 
in infrastructure \texttt{I}, at location \texttt{l}
if there exists a pair \texttt{(p,e)} in the local policy of \texttt{l}
(\texttt{delta I l} projects to the local policy) such that the action 
\texttt{a} is a member of the action set \texttt{e} and the policy 
predicate \texttt{p} holds for actor \texttt{h}.
\begin{ttbox}
enables I l h a \ttequiv \ttexists (p,e) \ttin delta I l. a \ttin e \ttand p h
\end{ttbox} 
%These are the main components to express the macro-level of infrastructures.
%for insider attacks.

%\subsection{Set up Isabelle Infrastructure Model}
%\subsection{System Model in Isabelle}
%\subsection{System Model}
% Describe the model of the IoT Healthcare system
%\subsection{SUCCESS Security and Privacy for IoT Healtcare}
%\label{sec:success}
% short summary of project and goals - could be part of some backgrond section
% or part of the intro to case study
%First, we introduce the application domain of transparent security and privacy of
%The  example of an IoT healthcare systems is from the CHIST-ERA project SUCCESS \cite{suc:16}
%on monitoring Alzheimer's patients. % (Section \ref{sec:success}). 
%Figure \ref{fig:iot} illustrates the system architecture where data collected by sensors 
%in the home or via a smart phone helps monitoring bio markers of the patient. The data 
%collection is in a cloud based server to enable hospitals (or scientific institutions) 
%to access the data which is controlled via the smart phone.
%\begin{figure}[h]
%\includegraphics[scale=.19]{gdpr_iot_success}
%\caption{IoT healthcare monitoring system for SUCCESS project}\label{fig:iot}
%\end{figure}
%We 
%omit the encoding of the \texttt{igraph} for the system architecture 
%shown in Section \ref{sec:app} in the Isabelle Infrastructure model 
%and instead 
%show how the policy is defined. 
%Given the graph of the infrastructure \texttt{G} 
The global policy is `only the patient and the doctor can access the data in the cloud':
%{\bf fixes} global_policy::[infrastructure, identity] \ttfun bool
%{\bf defines}  
\begin{ttbox}
 global_policy I a \ttequiv  a \ttnin hc_actors 
               \ttimp \ttneg(enables I cloud (Actor a) get)
\end{ttbox}

\subsection{Infrastructure State Transition}
The state transition relation uses the syntactic infix notation 
\texttt{I \ttrelI\, I'}  to denote that infrastructures 
\texttt{I} and \texttt{I'} are in this relation.
%Rules for the put, get, move, process and delete actions exist.
To give an impression of this definition, we show here just one of
several rules that defines the state transition for the action
get because this rule will be adapted in the process of refining the
system specification. Initially, this rule expresses that
an actor that resides at a location \texttt{l} (\texttt{h \ttatI\ l})
and is enabled by the local policy in this location to ``get'' can change the
state of that location to the string value \texttt{s} 
representing data stored in location \texttt{l'}.
\begin{ttbox}
 {\bf{get_data}}: G = graphI I \ttImp h \ttatI l \ttImp  
 l \ttin nodes G \ttImp l' \ttin nodes G \ttImp
 enables I l' (Actor h) get \ttImp s \ttin lgra G l' \ttImp
 I' = Infrastructure 
       (Lgraph (gra G)(agra G)(cgra G)
               (lgra G (l := lgra G l \ttcup \{s\})))
       (delta I) 
 \ttImp I \ttrel{n} I' 
\end{ttbox}
% ledgra G (n, (Actor a', as)) = L \ttImp l' \ttin L \ttImp 
%           (ledgra G (n, (Actor a', as)) := L \ttcup \{l\})
%move. The following inductive rule states that if an actor \texttt{h} 
%resides in a location \texttt{l} of the infrastructure graph \texttt{G} and
%a target's location \texttt{l'} local policy entitles this actor to the 
%move action, then the infrastructure \texttt{I} can transit into the infrastructure
%\texttt{I'} where \texttt{I'} is defined by an auxiliary function \texttt{move\_graph}
%(omitted here, for details see \cite{kam:18smc})
%\begin{ttbox}
%{\bf{move}}: 
%G = graphI I \ttImp h, h' \ttin actors_graph(graphI I) \ttImp
%l, l' \ttin nodes G \ttImp enables I l' (Actor h) move \ttImp
% I' = Infrastructure (move_graph_a a l l' 
%       (graphI I))(delta I)(tspace I)(lspace I) 
%\ttImp I \ttrelI I' 
%\end{ttbox}
Based on this state transition and the above defined\\
\texttt{hc\_scenario}, we define the first Kripke structure.
\begin{ttbox}
 hc_Kripke \ttequiv 
  Kripke \{ I. hc_scenario \ttrelIstar I \} \{hc_scenario\}
\end{ttbox}

\subsection{Attack: Eve can get data}
\label{sec:getatt}
How do we find attacks? The key is to use invalidation \cite{kp:14}
of the security property we want to achieve, here the global policy.
Since we consider a predicate transformer semantics, we use
sets of states to represent properties. 
%For example, the attack property
The invalidated global policy is given by the following set \texttt{shc}.
\begin{ttbox}
 shc \ttequiv \{x. \ttneg (global_policy x ''Eve'')\}
\end{ttbox}
The attack we are interested in is to see whether for the scenario
\begin{ttbox}
 hc\_scenario \ttequiv  Infrastructure ex_graph local_policies 
\end{ttbox}
from the initial state \texttt{Ihc \ttequiv \{hc\_scenario\}},
the critical state \texttt{sgdpr} can be reached,
that is, is there a valid attack \texttt{(Ihc,shc)}?

For the Kripke structure
\begin{ttbox}
 hc_Kripke \ttequiv Kripke \{ I. hc_scenario \ttrelIstar I \} Ihc
\end{ttbox}
we first derive a valid and-attack using the attack tree proof calculus.
\begin{ttbox}
\ttvdash [\ttcalN{(Ihc,HC)},\ttcalN{(HC,shc)}]\ttattand{\texttt{(Ihc,shc)}}
\end{ttbox}
The set \texttt{HC} is an intermediate state where \texttt{Eve} accesses the cloud.

The attack tree calculus \cite{kam:18b} exhibits that an attack is possible.
\begin{ttbox}
 hc_Kripke \ttvdash {\sf EF} shc
\end{ttbox}
We can simply apply the Correctness theorem \texttt{AT\_EF} to 
immediately prove this CTL statement. This application of the meta-theorem 
of Correctness of attack trees saves us proving the CTL formula tediously 
by exploring the state space in Isabelle proofs. Alternatively, we could use 
the generated code for the function \texttt{is\_attack\_tree} in Scala 
(see Section \ref{sec:at}) to check that a refined attack of the above is valid.

\section{Entering the Cycle}
\label{sec:apploop}
\subsection{First Refinement Iteration: Adding DLM Access Control}
%\subsection{Security and Privacy by Labeling Data}
\label{sec:label}
% Short summary of DLM
%\label{sec:ext}
The Decentralised Label Model (DLM) \cite{ml:98} allows
labelling data with owners and readers. We adopt it for our model.
% but propose
%extending labels with additional information, for example, purpose and retention 
%time, following the data protection requirements of the  GDPR. 
%In this section, we first present how we introduce DLM labeling into our 
%Isabelle Infrastructure model and how it can be used for privacy control. 
%%At the 
%%point where the semantics of eval actions is defined we can clearly identify the
%The difficulty when trying to enforce labels in distributed systems is that
%there is no control on what happens outside the scope of one unit of distribution. 
%We will therefore introduce a model of a distributed ledger that controls data in the
%system.
%This permits to subsequently show how these concepts of distributed labels and ledger
%can be employed to guarantee the consistent preservation of data protection in all parts 
%of a distributed system as requested by the GDPR.
%We specify the owner and the set of readers using the type \texttt{dlm}.
%\begin{ttbox}
%type dlm = actor \tttimes actor set
%\end{ttbox}
Labelled data is given by the type \texttt{dlm \tttimes\ data}
where \texttt{data} can be any data type. 
%Additional meta-data, like retention
%time and purpose, can be encoded as part of this type \texttt{data}. 
%We omit these details here for conciseness of the exposition.
%We may use the concept of erasure \cite{hs:08} to implement the latter.
% Basic definition used then in final step
%Using labeled data, we can now express the essence of Article 4
%Paragraph (1): 'personal data' means any information relating to an 
%identified or identifiable natural person ('data subject').
%Since we have a more constructive system view, we express this by
%defining the owner of a data item \texttt{d} of type \texttt{data \tttimes dlm}
%as the actor that is the first element in the pair that is the second 
%of the pair \texttt{d}.
%Then, we use this function to express the predicate ``owns''. 
%\begin{ttbox}
%  owner d \ttequiv fst(snd d)
%  owns G l a d \ttequiv owner d = a
%\end{ttbox}
%The introduction of a similar function for readers projecting the second element
%of a \texttt{dlm} label 
%\begin{ttbox}
%  readers d \ttequiv snd (snd d)
%\end{ttbox}
We provide functions \texttt{owns} and \texttt{readers} that
enable specifying when an actor may access a data item.
\begin{ttbox}
 has_access G l a d \ttequiv owns G l a d \ttor a \ttin readers d
\end{ttbox}
In the first refinement of the model in the RR-Cycle, we thus use labeled data 
to adapt the infrastructures. 
\subsubsection{Refinement Map}
\label{sec:refmapone}
Isabelle allows overloading of constant names, that is, the same name can be used
for different constants if these constants differ in their types or reside in 
different theories. We use the latter option and redefine the type 
\texttt{infrastructure} in a new theory \texttt{RRLoopTwo} (the original one was
called \texttt{RRLoopOne}). Included in that redefinition of the new type is
the redefinition of all involved constructors and projection functions.
Note that Isabelle's overloading permits the use of the same names.
To disambiguate these equally named constructors, we can make use of the 
Isabelle name spaces: \texttt{RRLoopOne.infrastructure} and \\
\texttt{RRLoopTwo.infrastructure} allow to reference the different types
of infrastructures and equally \texttt{RRLoopOne.gra} and 
\texttt{RRLoopTwo.gra}, for example, refer to the two \texttt{igraph}
projection functions. The extended names including the theory name
\texttt{RRLoopOne} or \texttt{RRLoopTwo} need only be used if the 
disambiguation is necessary (as for example below when we define the 
refinement map for this concrete first refinement step). However, 
as long as Isabelle can disambiguate the name from the context,
we can use the single names, for example, \texttt{infrastructure}
or \texttt{gra}.

In the refined model \texttt{RRLoopTwo}, the new type \\
\texttt{infrastructure} keeps now \texttt{dlm \tttimes\ data} 
instead of just data in the \texttt{igraph}.

Additionally as a preparation for defining the refinement, 
we need to define now a function from the new infrastructure type to 
the old one that projects out the data labels we have just introduced.
This is visible in the last input to the \texttt{Lgraph} constructor where
we map out the first data component \texttt{snd(RRLoopTwo.lgra (graphI I) l)}
of a \texttt{dlm \tttimes\ data} pair for each location \texttt{l}.
The function \texttt{fmap} is a "map" function for finite sets that we defined 
ourselves (see also Section \ref{sec:errors}).
\begin{ttbox}
 {\bf{definition}} refmap :: RRLoopTwo.infrastructure \ttfun 
                     RRLoopOne.infrastructure
 {\bf{where}} ref_map I = 
 RRLoopOne.Infrastructure 
        (RRLoopOne.Lgraph
           (RRLoopTwo.gra (graphI I))
           (RRLoopTwo.agra (graphI I))
           (RRLoopTwo.cgra (graphI I))
     (\ttlam l. fmap snd (RRLoopTwo.lgra (graphI I) l)))
\end{ttbox}
In the above expression, we deliberately put the theory names 
\texttt{RRLoopOne} and \texttt{RRLoopTwo} for all constructors and types
to enhance the understanding. 
In fact, this is only necessary for the type definition in the first line. 
For the constructors, e.g. \texttt{agra}, in the actual definition 
these can be omitted since Isabelle is capable of disambiguating
them from the context.
\subsubsection{Refined State Transition}
Also the state transition is now redefined for the refined
theory \texttt{RRLoopTwo} while keeping the same name and also 
overloading the infix syntax $\ttrel{n}$.
This first refinement iteration now implements access control in the
labeled data type but we also need to redefine the semantics of the
state transition.
The refined rule \texttt{get\_data} 
checks the labels for the data item stored in a location
\texttt{l'} and only gives access if -- in addition to get being enabled for an
actor \texttt{h} -- also this actor is among the readers or is the owner.
In this case, the data item including the label can be copied to
the location \texttt{l} where \texttt{h} resides.
\begin{ttbox}
 {\bf{get_data}}: G = graphI I \ttImp h \ttatI l \ttImp 
 l \ttin nodes G \ttImp l' \ttin nodes G \ttImp 
 enables I l (Actor h) get \ttImp 
 ((Actor h', hs), n) \ttin (lgra G l') \ttImp 
 Actor h \ttin hs \ttor h = h' \ttImp
 I' = Infrastructure 
        (Lgraph (gra G)(agra G)(cgra G)
                lgra G (l := lgra G l 
                       \ttcup \{((Actor h', hs), n)\}))
        (delta I) 
 \ttImp I \ttrel{n} I' 
\end{ttbox}

\subsubsection{Proof of Refinement}
We put those extension together by redefining a new Kripke structure 
\texttt{hc\_KripkeT}.
\begin{ttbox}
 hc_KripkeT \ttequiv 
  Kripke \{I. hc_scenarioT \ttrelIstar I\} \{hc_scenarioT\}
\end{ttbox}
In the above, we also use the redefinition of the involved infrastructure
states, for example, \texttt{hc\_scenarioT}, where the \texttt{T} stands
for \texttt{Two}. Note that these are definitions in a locale; they need 
to have different names since locales have a flat name space. 

However, these preparation pay off since we can finally apply our refinement 
theory from Section \ref{sec:transform} to prove 

\texttt{hc\_Kripke \ttmref{\texttt{refmap}} hc\_KripkeT}.

Moreover, we can use in addition the meta-theory about refinement
developed there: applying the theorem \texttt{strong\_mt'} 
allows to reduce this proof obligation to showing 
\begin{ttbox}
 ref_map \ttimg init hc_KripkeT \ttsubseteq init hc_Kripke \ttand
 (\ttforall s s'. (\ttexists s0 \ttin init K'. s0  \ttrelI s) \ttand 
               s \ttrelI s' \ttimp rmapT s \ttrelI rmapT s').
\end{ttbox}

\subsubsection{Attack: Eve can change labels}
\label{sec:atteval}
%The above ``get'' attack is still valid in the refined model but this
%does not matter any more since the global policy changes. 
%%Before providing the means to express this, we can however, already observe
%We cannot quite express the new policy yet before refining the model but
We can already observe another attack: Eve can also process data using the 
eval action at the cloud: we can prove there is a path ({\sf EF})
in the system leading to the corresponding attack state.
\begin{ttbox}
 hc_KripkeT \ttvdash 
     {\sf EF} \{I. enables I cloud (Actor ''Eve'') eval\}
\end{ttbox}
Once we have proved this CTL statement, we can use the Completeness theorems for the 
attack tree calculus (see Section \ref{sec:at}) and can thus derive that an attack 
exists:
Eve can tamper with the access control labels by processing labeled data.
%What we are interested in is to 
We need to prove privacy preservation, i.e. that labels are preserved.
As a countermeasure to this attack, the next iteration of the refinement cycle 
thus enforces label preserving functions.

\subsection{Second Iteration: Privacy Preservation}
\label{sec:rrtwo}
{\it The labels of data must not be changed by processing}.
%: we have identified this
%finally as an invariant (3) resulting from the GDPR in Section \ref{sec:gdpr}.
This invariant can be formalized in our Isabelle model by a type definition 
of functions on labeled data that preserve their labels.
\begin{ttbox}
{\bf{typedef}} label_fun = \{f :: dlm \tttimes data \ttfun dlm \tttimes data. 
                       \ttforall x. fst x = fst (f x)\}  
\end{ttbox}
We also define an additional function application operator \texttt{\ttupdownarrow}
on this new type. Then we can use this restricted function type to implicitly
specify that only functions preserving labels may be applied in the definition
of the system behaviour in the state transition rules.

This additional type definition \texttt{label\_fun} and its accompanying
operators build the core of the refined theory\\
\texttt{RRLoopThree} where
we also redefine the infrastructure type and corresponding operators and
projection functions.

\subsubsection{State Transition Refinement}
\label{sec:stref3}
The crucial point for this refinement to \texttt{RRLoopThree} is 
that the state transition changes to incorporate the new restrictions on
label processing. The rule for eval now enforces the use of labelled functions.

\paragraph{The process rule} This rule prescribes how data within the
infrastructure may be processed. It imposes that only privacy preserving functions
may be applied to data (see Section \ref{sec:ppfun}). This is achieved by
using the application operator \texttt{\ttupdownarrow} because it enforces
the variable \texttt{f} to be of type \texttt{label\_fun}. % which preserves the label. 
The existing data item \texttt{((Actor a', as), n)} is replaced
by \texttt{f \ttupdownarrow ((Actor a', as), n)} while preserving the
label owing to the properties of type \texttt{label\_fun}.
Clearly, the actor needs to be \texttt{eval} enabled in his
location where also the data must reside.
%The update changes the ledger's {\it domain} by re-assigning -- again using update --
%the location set \texttt{L} to the new input \texttt{(f \ttupdownarrow (n, (Actor a', as)))}
%of the ledger function \texttt{ledgra G}. 
%Also, the old value of the data item \texttt{(n, (Actor a', as)})
%is deleted by assigning it to the empty set \texttt{\{\}} to preserve the 
%invariant of the legder type. % (and thus the resulting consistency, see Section \ref{sec:cons}).
%Note, that this semantics of process changes the data on processing consistently in all parts
%of the distributed system (see resulting consistency property in Section \ref{sec:guar}).

\begin{ttbox}
{\bf{process}}: G = graphI I \ttImp h \ttatI l \ttImp 
l \ttin nodes G \ttImp enables I l (Actor h) eval \ttImp 
((Actor h', hs), n) \ttin lgra G l \ttImp 
Actor h \ttin hs \ttor h = h' \ttImp 
I' = Infrastructure 
      (Lgraph (gra G)(agra G)(cgra G)
       ((lgra G)(l := lgra G l  - \{(y, x). x = n\}
        \ttcup \{(f::label_fun)\ttupdownarrow((Actor h', hs), n)\}))))
       (delta I)
\ttImp I \ttrel{n} I'
\end{ttbox}
\paragraph{Processing preserves privacy}
\label{sec:privpres}
Furthermore, we can prove now that only entitled users (owners and readers) 
can access data: privacy is preserved by the use of label preserving functions.
We can prove that processing preserves ownership for all paths 
globally (expressed using the CTL quantifier \texttt{\sf AG}).
That is, in all states of the Kripke structure and all locations of 
the infrastructure graph we have that the ownership in the initial
state \texttt{hc\_scenario} will persist.
%% used in esorics paper since referred to:
%%The Kripke structure \texttt{gdpr\_Kripke} is defined over
%%an intitial infrastructure \texttt{gdpr\_scenario} and the set
%%\texttt{gdpr\_states} of all infrastructure states reachable from there. 
%%\begin{ttbox}
%% gdpr_Kripke \ttequiv Kripke gdpr_states \{{gdpr_scenario}\}
%%\end{ttbox}
%Note, that it would not be possible to express such a set
%using a universally quantified formula within a temporal operator
%when using Modelcheckers since they only allow propositional logic 
%within states.
%This generalisation is only possible since we use Higher Order 
%Logic. % (see also Section \ref{sec:concl}).
\begin{ttbox}
{\bf{theorem}} priv_pres: h \ttin hc_actors \ttImp 
  l \ttin hc_locations \ttImp 
  owns (Igraph hc_scenario) l (Actor h) d \ttImp
  hc_KripkeR \ttvdash {\sf AG} {x. \ttforall l \ttin hc_locations. 
                     owns (Igraph x) l (Actor h) d }  
\end{ttbox}
%While this property has already been proveable with 
%distributed labels and privacy preserving functions alone \cite{kam:18a}, 
%the following global consistency is only proveable thanks to the ledger.
\subsubsection{Refinement Map}
as in the previous step, we define the refined Kripke structure 
\texttt{hc\_KripkeR} now for the refined theory \texttt{RRLoopThree}
and the redefined infrastructure states, here for example,
\texttt{hc\_scenarioR} as initial state of the state transition.
\begin{ttbox}
 hc_KripkeR \ttequiv 
  Kripke \{ I. hc_scenarioR \ttrelIstar I \} \{hc_scenarioR\}
\end{ttbox}
Also a new datatype map \texttt{ref\_mapR} mapping 
the infrastructure type of \texttt{RRLoopThree} to  that of
\texttt{RRLoopTwo} is defined. Note that here we only need to 
re-embed the constituents of the infrastructure of \texttt{RRLoopThree}
with the corresponding constructors of \texttt{RRLoopTwo} within the 
definition. In the previous refinement map we needed to map out
the first element of each \texttt{dlm \tttimes\ data} pair for each
location. Now the actual structure of the labeled data is very similar
but the function type changes to labeled functions. This happens automatically
by the re-embedding.
\begin{ttbox}
{\bf{definition}} refmapR :: RRLoopThree.infrastructure \ttfun 
                    RRLoopTwo.infrastructure
{\bf{where}} ref_mapR I = RRLoopTwo.Infrastructure 
                     (RRLoopTwo.Lgraph
                       (RRLoopTwo.gra (graphI I))
                       (RRLoopTwo.agra (graphI I))
                       (RRLoopTwo.cgra (graphI I))
                       (RRLoopTwo.lgra (graphI I)))
\end{ttbox}
Applying our refinement theory we now prove 

\texttt{hc\_KripkeT \ttmref{\texttt{ref\_mapR}} hc\_KripkeR}.

It is important to note that by the additional meta-theorem \texttt{prop\_pres}
the properties proved for \texttt{hc\_Kripke} and\\
\texttt{hc\_KripkeT} carry
over to \texttt{hc\_KripkeR}. 

\subsubsection{Attack: Eve can simply put data}
\label{sec:attput}
When trying to prove a theorem to express that different occurrences of the
same data in the system must have the same labels, we fail.
The reason for this is the following attack.
\begin{ttbox}
 hc_KripkeR \ttvdash 
  {\sf EF} {I. enables I cloud (Actor ''Eve'') put}
\end{ttbox}
Eve could learn the data by other means than using the privacy preserving
functions and using the action put to enter 
that data as new data to the system labelled as her own data. 
%We skip the corresponding 
%attack tree here and immediately consider 
As a countermeasure, we need a concept to guarantee consistency across the system: 
blockchain.

\subsection{Third RR-Cycle Iteration: Blockchain Consistency}
\label{sec:ppfun}
One major achievement of a blockchain is that it acts like a distributed ledger, that
is, a global accounting book.
A distributed ledger is a unique consistent transcript keeping track of
protected data across a distributed system. 
%It is commonly implemented by 
%mechanisms like a blockchain that logs all transactions on data between 
%different units of distribution. 
In our application, the ledger must mainly keep track of where the data resides
for any labelled data item.
We formalize a ledger thus as a type of functions that maps a labelled
data item to a set of locations. In this type, we further constrain each data
to have at most one valid data label of type \texttt{dlm}. This is achieved by
stating that there exists a unique (\texttt{\ttexists{!}}) label l for which
the location set \texttt{ld(l, d)} assigned to by the ledger is {\it not} empty
-- unless it is empty for all labels for \texttt{d}.
% old idea
%{\bf typedef} ledger\_fun = \{ lf ::  data \ttfun dlm. inj lf \}
\begin{ttbox}
{\bf{typedef}} ledger = \{ ld :: dlm \tttimes data \ttfun location set.
                    \ttforall d. (\ttforall l. ld (l, d) = \{\}) \ttor 
                         (\ttexists{!} l. ld (l, d) \ttneq \{\}) \}
\end{ttbox}
%The type location represents the locations in infrastructures as presented next.
%lifts this type to a flat domain, that is,
%it adds an undefined element (or bottom) thereby allowing a ledger to be 
%undefined for certain inputs. This effectively turns the ledger into a 
%partial functions since it might not contain certain data inputs at all.
The addition of \texttt{set} 
makes the range of the ledger a set of sets of locations which allows for none (empty set)
or a number of locations to be assigned to a data item.

\subsubsection{Ledger enables Data Protecting State Transition}
%% Explain how the semantics implements the GDPR rule
%The abstract state transition provided in the underlying Kripke structure
%theory is instantiated in the GDPR infrastructure model by an inductive definition 
%of a state transition relation \texttt{\ttrel{n}} over infrastructures. 
%%A set of inductive rules defines this transition relation \texttt{\ttrel{n}} 
%%relative to characteristics of the current state. These characteristics can exploit
%%the information encoded into the infrastructure as well as the enables predicate
%%to express how the next infrastructure state evolves from the current one.
The set of rules for defining the state transition of infrastructures needs to be 
adapted to the refined model.
The %GDPR articles 
refinement by a ledger
is incarnated into the system specification to guarantee consistency across distributed 
units. The state transition rules have to be adapted yet again but also the type 
\texttt{dlm} needs to be refined replacing actors by identities since otherwise 
the uniqueness of the label imposed in the ledger typedef cannot be proved for 
actors.
The abstract models intentionally did not stipulate Actor to be injective to allow 
for insider attacks -- now the ledger enforces the use of identities rather than 
actor ``roles''.

\subsubsection{State Transition Refinement}
We illustrate the changes of this refinement step again on the rule for 
\texttt{get} first. Since now the model is fairly complete, we finally also show
the other rules.
%for reasons of space the Isabelle source code \cite{kam:18smc} for details).

\paragraph
{The get data rule} 
%resembles the put data rule in 
%many parts. However, here an actor \texttt{a} accesses data in a remote
%location \texttt{l'} and adds it to the data in his current location  
%\texttt{l}. This copying of data is only permitted 
This rule now requires that %if the current location
%\texttt{l'}  of the data enables \texttt{a} to \texttt{get} and if the list of readers
%\texttt{as} in the label \texttt{(Actor a', as)} of a data item \texttt{n} contains the entry 
%\texttt{Actor a}. 
%%Different to the put rule, this rule preserves the first component 
%%\texttt{fst(lgra G l)} of the state of location \texttt{l}.
the ledger be updated by noting that the data item also resides
in the new location \texttt{l}. This is achieved by unifying the existing
set of locations \texttt{L} for this data item with the new location \texttt{l}.
The existing set of locations \texttt{L} is simply retrieved by applying the
ledger \texttt{ledgra G} to the data item \texttt{n} and its label \texttt{(h', hs)}. 
The update of the ledger at the position \texttt{ledgra G ((h', hs), n)} of this data item 
uses the operator \texttt{:=} to change the ledger to contain the new list of 
locations \texttt{L \ttcup\ \{l\}}.
\begin{ttbox}
 {\bf{get_data}}: G = graphI I \ttImp h \ttatI l \ttImp 
 l \ttin nodes G \ttImp l' \ttin nodes G \ttImp 
 enables I l' (Actor h) get \ttImp 
 Actor h \ttin hs \ttor h = h' \ttImp 
 ledgra G (n, (Actor h', hs)) = L \ttImp l' \ttin L \ttImp 
 I' = Infrastructure 
       (Lgraph (gra G)(agra G)(cgra G)(lgra G)
            (ledgra G ((h', hs), n) := L \ttcup \{l\})
       (delta I) 
 \ttImp I \ttrel{n} I' 
\end{ttbox}

\paragraph{The put data rule} It assumes an 
actor \texttt{h} residing at a location \texttt{l} in the infrastructure
graph \texttt{G} and being enabled the \texttt{put} action. If infrastructure 
state \texttt{I} fulfils those preconditions, the next state \texttt{I'} can
be constructed from the current state by adding the data item 
\texttt{n} with label \texttt{(h, hs)} at location \texttt{l}. 
The addition 
is given by updating (using \texttt{:=}) the existing ledger \texttt{ledgra G}.
The ledger is set for this labelled data item \texttt{(n, (h, hs))} initially 
as the singleton set \texttt{\{l\}} containing just this location.
%%data storage \texttt{lgra G l} 
%%at location \texttt{l} with the set union of its second element and the singleton
%%set \texttt{\{((Actor a, as), n)\}}. 
Note that the first component \texttt{h} marks the owner of this data item 
as \texttt{h}. The other components are the reader list \texttt{hs}, and the actual 
data \texttt{n}.%%, as well
%%as the state of the location \texttt{s} %have no restrictions;. They 
%%can be instantiated freely within the limitations given by the Isabelle types.
\begin{ttbox}
 {\bf{put}}: G = graphI I \ttImp h \ttatI l \ttImp 
 enables I l (Actor h) put \ttImp
 I' = Infrastructure 
       (Lgraph (gra G)(agra G)(cgra G)(lgra G)
               (ledgra G ((Actor h, hs), n) := {l})) 
       (delta I) 
 \ttImp I \ttrel{n} I' 
\end{ttbox}
\paragraph{The process rule} This rule is now simplified by use of the ledger.
%%\subsubsection{The process rule} prescribes how data within the
%%infrastructure may be processed. It imposes that only privacy preserving functions
%%may be applied to data (see Section \ref{sec:ppfun}). This is achieved by
%%using the application operator \texttt{\ttupdownarrow} because it enforces
%%the variable \texttt{f} to be of type \texttt{label\_fun}. % which preserves the label. 
%%The existing data item \texttt{(n, (Actor a', as))} is replaced
%%by \texttt{f \ttupdownarrow (n, (Actor a', as))} while preserving the
%%label owing to the properties of type \texttt{label\_fun}.
%%Clearly, the actor needs to be \texttt{eval} enabled in his
%%location where also the data must reside.
The update changes the ledger's {\it domain} by re-assigning -- again using update --
the location set \texttt{L} to the new input \texttt{(f \ttupdownarrow ((a', as),n))}
of the ledger function \texttt{ledgra G}. 
First, the old value of the data item \texttt{((a', as), n})
is deleted by assigning it to the empty set \texttt{\{\}} to preserve the 
invariant of the ledger type. % (and thus the resulting consistency, see Section \ref{sec:cons}).
Note, that this semantics of process changes the data on processing consistently in all parts
of the distributed system (see resulting consistency property in Section \ref{sec:guar}).
%%The additional condition : validity of f 
\begin{ttbox}
 {\bf{process}}: G = graphI I \ttImp a \ttatI l \ttImp 
 enables I l (Actor a) eval \ttImp
 a \ttin as \ttor a = a' \ttImp ledgra G ((a', as), n) = L \ttImp
 I' = Infrastructure 
       (Lgraph (gra G)(agra G)(cgra G)(lgra G)
               (ledgra G ((a', as), n) := \{\})
                        (f \ttupdownarrow((a', as),n)):= L)                  
       (delta I) 
 \ttImp I \ttrel{n} I' 
\end{ttbox}

\paragraph{The delete rule} The owner of the data may delete his or her data from {\it all} 
locations in the infrastructure graph.
Note that, different to the previous rules, here are no preconditions on the 
location of the actor nor the location of the data other than that they are
in the infrastructure graph. Neither is there any requested enabledness of 
actions imposed on the actor. That is, the owner can delete his data anywhere.
Also note, how the use of the ledger simplifies the deletion of data throughout the
system: it suffices to update the {\it ledger} to the empty set; automatically 
the data is deleted everywhere.
\begin{ttbox}
 {\bf{del_data}}: G = graphI I \ttImp a \ttin actors G \ttImp 
 l \ttin nodes G \ttImp l \ttin L \ttImp
 ledgra G ((a', as), n) = L \ttImp
 I' = Infrastructure 
       (Lgraph (gra G)(agra G)(cgra G)(lgra G)
               (ledgra G ((a', as), n) := \{\}))
 \ttImp I \ttrel{n} I' 
\end{ttbox}

\paragraph{The move rule} This rule completes the set of inductive rule
defining the semantics of the state transition relation \texttt{\ttrel{n}}.
This inductive rule states that if an actor \texttt{h} 
resides in a location \texttt{l} of the infrastructure graph \texttt{G} and
a target's location \texttt{l'} local policy entitles this actor to the 
move action, then the infrastructure \texttt{I} can transit into the infrastructure
\texttt{I'} where \texttt{I'} is defined by an auxiliary function \texttt{move\_graph}
(omitted here, for details see \cite{kam:18smc})
\begin{ttbox}
 {\bf{move}}: G = graphI I \ttImp 
 h \ttin actors_graph(graphI I) \ttImp
 h' \ttin actors_graph(graphI I) \ttImp l \ttin nodes G \ttImp  
 l' \ttin nodes G \ttImp enables I l' (Actor h) move \ttImp
 I' = Infrastructure 
        (move_graph_a a l l' (graphI I))
        (delta I)
 \ttImp I \ttrelI I' 
\end{ttbox}

%\subsection{Refinement Map}
%\label{sec:isaapp}
\subsubsection{Refinement Map}

In the extended infrastructure of the refined system the infrastructure graph needs to be extended
by the ledger. The resulting \texttt{infrastructure} in the refined theory
\texttt{RRLoopFour} thus contains a ledger. 
So, the refinement map needs to transform the ledger in the infrastructure 
graph into a map from locations to sets of labeled data.
\begin{ttbox}
 {\bf{definition}} refmapF :: RRLoopFour.infrastructure \ttfun 
                        RRLoopThree.infrastructure
 {\bf{where}} 
 ref_mapF I = RRLoopThree.Infrastructure 
               (RRLoopThree.Lgraph
               (RRLoopThree.gra (graphI I))
               (RRLoopThree.agra (graphI I))
               (RRLoopThree.cgra (graphI I))
               (ledger_to_loc (ledgra (graphI I)))
\end{ttbox}
The projection \texttt{ledgra} just maps out the ledger but the 
auxiliary function \texttt{ledger\_to\_loc} performs the main data
type transformation defined by the following functions.
\begin{ttbox}
 dlm_to_dlm \ttequiv (\ttlam ((s :: string), (sl :: string set)). 
                   (Actor s, fmap Actor sl))
 data_trans  \ttequiv 
   (\ttlam (l :: (string \tttimes string set),d :: string). 
      (dlm_to_dlm l, d))
 ledger_to_loc ld l \ttequiv 
   if l \ttin {\sf U} range(Rep_ledger ld) 
   then fmap data_trans \{dl. l \ttin (ld dl)\} else \{\}
\end{ttbox}
The function \texttt{Rep\_ledger} is the injection from elements of the
ledger type into the set defining the type. It is automatically created by Isabelle
from the type definition.

To make the refinement proofs feasible, it is necessary to provide a set
of rather technical lemmas to support the use of this transformation within the
refinement map. For details see \cite{kam:18smc}.
A central lemma is clearly the uniqueness of the data given by the labels.
\begin{ttbox}
 {\bf{lemma}} ledger_to_loc_data_unique: 
 Rep\_ledger ld (dl,d) \ttneq \{\} \ttImp 
 Rep\_ledger ld (dl',d) \ttneq \{\} \ttImp dl = dl'
\end{ttbox}
Central as well is a transformation lemma.
\begin{ttbox}
 {\bf{lemma}} ledgra_ledger_to_loc:
 finite\{dl::(char list\tttimes{char list set})\tttimes{char list}. 
          l \ttin Rep\_ledger (ledgra G) dl\} \ttImp
 l \ttin (ledgra G ((a, as), n)) \ttImp 
 ((Actor a, fmap Actor as), n) \ttin 
        ledger_to_loc(ledgra G) l
\end{ttbox}
As before we show
\texttt{hc\_KripkeR \ttmref{\texttt{ref\_mapF}} hc\_KripkeF}
for the corresponding models.
The main change between those infrastructure models is due to the use of the ledger.
It is visible in the infrastructure graph where an additional component ,
here \texttt{ex\_ledger} appears.
\begin{ttbox}
ex_graph \ttequiv Lgraph 
 \{(home, cloud), (sphone, cloud), (cloud,hospital)\}
    (\ttlam x. if x = home then \{''Patient''\} else 
       (if x = hospital then \{''Doctor''\} else \{\})) 
       ex_creds ex_locs ex_ledger
\end{ttbox}
%The data and its privacy access control definition is given by
This parameter \texttt{ex\_ledger} specifies in our running example
that the data "42", for example, 
some bio marker's value, is owned by the patient and can be read by the doctor
and is currently only contained in location \texttt{cloud}.
\begin{ttbox}
ex_ledger \ttequiv (\ttlam (l, d).  
  if d = ''42'' \ttand l = (''Patient'',\{''Doctor''\}) 
  then \{cloud\} else \{\})
\end{ttbox}

\subsubsection{Ledger Guarantees Consistent Data Ownership}
\label{sec:guar}
We can now prove that data
protection is consistent across the infrastructure. If in any two locations
the same data item \texttt{n} resides, then the labeling must be the same.
That is, the owner and set of readers are identical.
\begin{ttbox}
{\bf{theorem}} Ledger_con: h \ttin hc_actors \ttImp 
h' \ttin hc_actors \ttImp  
l \ttin hc_locations \ttImp l' \ttin hc_locations \ttImp 
l \ttin ledgra G ((h, hs), n) \ttImp 
l' \ttin ledgra G ((h', hs'), n) \ttImp
(h, hs) = (h', hs')
\end{ttbox}
This property immediately follows from the invariant property of the
type definition of the type ledger (see Section \ref{sec:ppfun}) and privacy 
preservation given by the label function type (see Section \ref{sec:label}).
This means that the corresponding interactive proofs that we have to provide to 
Isabelle are straightforward and largely supported by its automated tactics (see
the Isabelle source code %\cite{kam:18smc} 
for details).

%Exploiting the invariant properties of the ledger given in the type
%definition 

\subsection{Attack and Fourth RR-Cycle: Eve can overwrite blockchain}
\label{sec:rrfour}
Despite the above proved theorem, there is yet another aspect -- as usual 
outside the model -- that leads to an attack. 
In the abstract specification of a ledger, we have omitted the implementation
of a blockchain.
We could have a centrally controlled blockchain in which one part signs the entire
blockchain to guarantee consistency.
Eve could be an insider impersonating the blockchain controller. 
In that case, she could just overwrite the entry made by Bob and add his data
as her own.
Formally, we can re-use the put attack of the previous level 
using the rule put above to overwrite Bob's entry by Eve's.

As a refinement for the RR-Cycle, we need to consider a consensus algorithm, like
Nakamoto's used in Bitcoin, between the participants in the distributed system to 
chose a different leader for each blockchain commitment to avoid the attack.
%The detail of this refinement of our model are beyond the scope of this paper. 
Adding a refinement with a Nakamoto consensus to our model is possible but 
rather complex. However, we can simply specify the effect of
this refinement in the system specification by adding  
\begin{ttbox}
\ttforall a as. ledgra G ((Actor a, as), n) = \{\}
\end{ttbox} 
as a precondition
to the rule put, that is, the data item must not yet be assigned to anyone 
in the ledger in order to allow a put action.

\subsection{Evaluation and Detecting Design Errors}
\label{sec:errors}
To give a a rough estimate of the formalisation and proof effort of the 
application of the Refinement-Risk cycle to the IoT healthcare application
provided in this paper: each of the 8 files (four pairs of files: one for the semantics 
and the other for the example infrastructure) has between 200 and 800 lines of 
Isabelle code: definitions and mostly proof script lines.

Clearly, an important motivation for going through this rather tedious process of 
formally refining a system specification in this framework is the property preservation that
we have established as a meta-theorem on the refinement in 
Section \ref{sec:transform}.
It allows us to preserve once gained security and privacy properties and increasingly 
make the specification more secure. 

The effort to do the refinement proofs is rather high: the proofs of refinement
in each level are up to 400 lines of Isabelle code and sometimes necessitated
proving additional lemmas about the new operators, for example, for label preserving 
functions and the ledger type.

As mentioned in the introduction and repeatedly throughout the paper, one main 
advantage of the formal security refinement approach presented in this paper is that
it filters out errors that are easily made in the stepwise design. 
We have found and corrected a number of small errors, like inconsistencies
of the premises in the rules for the state transition at the four different levels of
model abstraction. For example, the patient data was positioned in the cloud 
in \texttt{RRLoopThree} and at \texttt{home} in \texttt{RRLoopFour}.
The formal refinement with \texttt{\ttmref{\texttt{refmap}}} forces out these
errors immediately. The simple ones, like the former example, are easy to fix but 
others require more work to understand them, find a solution and provide the necessary
lemmas to then prove the refinement. The more subtle ones are sometimes harder to fix
like the following example shows.
%Here are the two main errors we found.
%However, the more subtle ones are not so easily spotted.

%\subsubsection{DLM Label elimination}
When we introduce the DLM labels in the first iteration, the corresponding
refinement map is based on the function \texttt{refmap} mapping the refined 
data type to the abstract data type. In this first map, we need to eliminate the
data label. So, we simply apply the function \texttt{snd} to all data items of 
type \texttt{dlm \tttimes data} to map to type \texttt{data}; formally
applying \texttt{fmap snd} (see Section \ref{sec:refmapone}).

In the earlier version of the IoT case study \cite{kam:19a}, the rule for delete 
uses set difference \texttt{-} to delete the labeled data item in \texttt{RRLoopTwo}.
\begin{ttbox}
{\bf{del_data}}: G = graphI I \ttImp h \ttin actors_graph  G \ttImp
l \ttin nodes G \ttImp ((Actor h, hs), n) \ttin lgra G l \ttImp
I' = Infrastructure 
      (Lgraph (gra G)(agra G)(cgra G)(lgra G)
              ((lgra G)(l := (lgra G l) - 
                       \{((Actor h, hs), n)\}))
      (delta I) 
\ttImp I \ttrel{n} I' 
\end{ttbox}
The subtle design error manifests itself when we try to prove that the semantics
using the above rule is a refinement of the abstract model \texttt{hcKripkeOne}. 
This semantics still allows the data item \texttt{n}
to occur with two different labels say, \texttt{(Actor h, hs) \ttneq\ (Actor h', hs')}
(pre-ledger model). We may have two similar traces where deletion appears 
once on \texttt{((Actor h, hs), n)} and once on \texttt{((Actor h', hs'), n)}.
The refinement map maps both traces to one trace in the abstract model \texttt{hcKripkeOne}.
In the abstract trace, after the deletion, the state does not contain the data item
\texttt{n} any more, while in both refined traces one copy of the data item
(with mutually different labels) prevails. Both are insecure states that must not implement
the abstract specification: a users data that is believed to be eradicated is still 
in the data base potentially with another label of an attacker. So, for privacy enforcement
it is absolutely crucial to avoid such design errors. 

This design error can be eradicated by making sure that a deletion operation actually
deletes all copies of the data item.
\begin{ttbox}
{\bf{del_data'}}: G = graphI I \ttImp h \ttin actors_graph  G \ttImp
l \ttin nodes G \ttImp ((Actor h, hs), n) \ttin lgra G l \ttImp
I' = Infrastructure 
      (Lgraph (gra G)(agra G)(cgra G)(lgra G)
              ((lgra G)(l := (lgra G l) - 
                        \{(y, x). x = n \}))
      (delta I) 
\ttImp I \ttrel{n} I' 
\end{ttbox}
The same problem occurs in the process rule but the same solution applies (Section
\ref{sec:stref3} shows the fixed rule).
Once this solution has been found, we need to prove that this fixed semantics preserves
traces now. A core lemma we need to prove to this end is the following.
\begin{ttbox}
{\bf{lemma}} fmap_lem_del_set: finite S \ttImp  
\ttforall n {\ttin} S. 
 fmap f (S - \{y. f y = f n\}) = (fmap f S) - \{f n\}
\end{ttbox}

%\subsubsection{Blockchain}
%the informal presentation of the RRLoop \cite{kam:19a}, we 

\section{Conclusion and Related Work}
\label{sec:concl}
In this paper, we have presented a formal integrated framework for a Refinement-Risk-Cycle 
that interleaves formal system specification with attack tree analysis by a 
refinement based on refinement. 
Thereby, formally proved engineering of the security of a system becomes possible.
The method is particularly useful for IoT systems since it allows modeling 
physical as well as logical realities.
We have illustrated this process on an IoT healthcare example running four 
iterations adding access control, privacy preservation, and a ledger for 
global consistency. Framework and casestudy are fully formalised and proved in Isabelle.

%\subsection{Discussion and Related Work}
Formal system specification refinement has been investigated for some time
initially for system refinement in the specification language Z \cite{hhs:86}
but a dedicated security refinement has not been formalised for some 
time \cite{mor:09}. 
%Only later, 
The idea to refine a system specification for security has been already
addressed in B \cite{bcmjk:07, sp:07}.
The former combines the refinement of B with system security policies given in 
Organisation based Access Control (OrBAC) and presents a generic example of a 
system development.
While B is supported by its own tool Atelier B, it does not provide a formalisation 
in a theorem prover unlike our integration which supports dedicated security concepts 
like attack trees and enables useful meta-theory over the integration.
The paper \cite{sp:07} looks at attacks within the B framework but it aims at 
designing a monitor that catches actions forbidden by the policy not on using these
attacks to refine the system specification.
Dynamic risk assessment using attack formalism, like attack graphs, has recently 
found great attention, e.g. \cite{gdmgampd:18}. However, usually,
the focus of the process lies on attack generation and response planning while we
address the design of secure systems. Rather than incident response, we intend to 
use early analysis of system specification to provide a development of secure systems.
This includes physical infrastructure, like IoT system architecture, as well as 
organisational policies with actors.

While the additional consideration of structural refinement in our process of security 
refinement greatly generalises the classical concepts of trace refinement, the latter
has been designed for safety properties. These are properties that hold along 
execution paths of a system.
This is known to be insufficient in general for security properties:
a security property often 
has to do with implicit information flows that may lead to an attacker learning some 
confidential information by observing various runs of a system over time thereby
noticing differences in the outcomes.  McLean has already shown in his seminal 
paper \cite{mcl:94} that for these kind of implicit information
flow properties, it is necessary to consider a security property as a 
{\it set of set} of traces rather than a set of traces, leading on to notions like
noninterference that have been formalised in Isabelle, e.g.,
\cite{DBLP:conf/cpp/0001HN13}.
However, we argue that our way of modeling systems
and their execution using a layered model of CTL and Kripke structures underneath
the actual infrastructure model including actors allows a more fine-grained view.
As we see in the example application, the explicit modeling of actors
allows reasoning about specific attackers as actors. Thereby, rather
than trying to establish security properties at the very basis of state based
system modeling, we propose to consider the analysis of information flows and 
their observability by certain actors at the level of modeling the actual
infrastructure. 
When actors are actually part of the model, it is more natural to 
add notions of implicit information flow and then postulating the reachability of 
security critical states -- in which an actor has learned some confidential information --
as a state. A trace based safety analysis of whether those states are reachable then 
results in the same analysis as a classical noninterference analysis -- our layered
model just clarifies the boundaries at a finer scale.

The use of a distributed ledger, also known as a blockchain, is new for
formal system specification and verification. There are currently many attempts to
formalize blockchains but most of them are very close to technical implementations,
e.g. \cite{kmswp:16}, thus obliterating the possibility to provide clear specification 
of legal requirements as is possible in the Isabelle Infrastructure framework and has been
illustrated on GDPR requirements \cite{kam:18a}.
Moreover, to our knowledge, none of these formal models has been produced in Isabelle or 
similar Higher Order Logic tools until very recently \cite{mar:19}, where Marmsoler
addresses the definition of an Isabelle framework for the verification of dynamic system architectures
for blockchain. Our formalization uses a generic notion of a ledger that 
may simply control consistency in the distributed application. 
This is the way forward 
because it enables the minimal expression of the crucial properties of a ledger. This 
minimal expression may not only be used as a basis for conformance proofs of more refined 
technical models of a ledger, like a blockchain, but also provides the crucial invariant
properties for a ledger.
We have used label preserving functions in our infrastructure model. They guarantee 
part of the data protection consistency when processing data. 
An interesting next step is to model smart contracts. We believe this to be a particularly 
rewarding extension because our model permits the expression of locality and policy based 
behaviour rules which naturally lends itself to allow modeling dependant action sequences
using logical preconditions. 

%Risk assessment loops exist for secure systems, e.g. \cite{gdmgampd:18}. There
%the process generates attacks in order to plan incident responses. By contrast, we
%use the risk assessment to improve the design of secure systems by refinement for
%system specification development of secure systems from formal specifications.
%This includes physical infrastructure, like IoT system architecture, as well as 
%organisational policies with actors.
Research on risk assessment and attack trees has recently increased using formal
approaches including verification, e.g., \cite{anp:16,apk:17,BCLQ18} but not in Higher Order Logic. With respect to system development the focus is often on the generation of 
the attack tree not the system, e.g. \cite{vnn:14}. In \cite{DBLP:conf/csfw/AudinotPK18}, the authors build a foundation for system based attack trees but do not mechanise it in
a theorem prover. 
Model transformation for attack trees has been addressed in 
\cite{DBLP:conf/fase/0012SRYHBRS18} as a practical tool to translate between
different frameworks but not to reason about refinement.
Data refinement has been addressed in Isabelle \cite{DBLP:conf/itp/HaftmannKKN13} but not
with respect to security engineering.

The relationship between Higher Order logic and Modelchecking has been first explored by
Kobayashi (see \cite{DBLP:journals/jacm/Kobayashi13} for a paper subsuming previous results).
Modelchecking has been realized as well in Isabelle 
\cite{DBLP:conf/cav/EsparzaLNNSS13} but we 
use the different formalisation of CTL \cite{kam:16b}.
Developing secure systems using Isabelle has been done using the formalisation of noninterference
to develop an online conference system \cite{klpbb:14}.

%\subsection{Outlook}
The novelty of our approach is to integrate formally refinement
with the risk assessment by attack trees into a constructive security refinement 
process. %This approach clearly demands a certain level of familiarity with logical 
%specification. However, 
Abstract system specifications can be 
provably refined and finally code can be extracted to major programming languages, e.g.
Scala.
%The larger aim of providing a logical framework for the design of actor based
%edge network systems is to channel the power of machine learning techniques:
%the logical expression and stepwise security refinement of system policies is
%a natural counter-part that when added can control smart computing within edge networks.

%% Acknowledgments
%\begin{acks}                            %% acks environment is optional
                                        %% contents suppressed with 'anonymous'
  %% Commands \grantsponsor{<sponsorID>}{<name>}{<url>} and
  %% \grantnum[<url>]{<sponsorID>}{<number>} should be used to
  %% acknowledge financial support and will be used by metadata
  %% extraction tools.
  This material is based upon work supported by the
  ERA-NET {CHIST-ERA}{http://dx.doi.org/10.13039/100000001} under Grant
  No. {102112}.  Any opinions, findings, and
  conclusions or recommendations expressed in this material are those
  of the author and do not necessarily reflect the views of the
  European Union.
%\end{acks}

%% Bibliography
\bibliographystyle{abbrv}
\bibliography{insider}

\begin{thebibliography}{10}

\bibitem{anp:16}
Z.~Aslanyan, F.~Nielson, and D.~Parker.
\newblock Quantitative verification and synthesis of attack-defence scenarios.
\newblock In {\em 29th IEEE Computer Security Foundations Symposium, CSF'16},
  2016.

\bibitem{apk:17}
M.~Audinot, S.~Pinchinat, and B.~Kordy.
\newblock Is my attack tree correct?
\newblock In {\em 22nd European Symposium on Research in Computer Security,
  ESORICS'2017}, volume 10492 of {\em LNCS}, pages 83--102. Springer, 2017.

\bibitem{DBLP:conf/csfw/AudinotPK18}
M.~Audinot, S.~Pinchinat, and B.~Kordy.
\newblock Guided design of attack trees: {A} system-based approach.
\newblock In {\em 31st {IEEE} Computer Security Foundations Symposium, {CSF}
  2018, Oxford, United Kingdom, July 9-12, 2018}, pages 61--75. {IEEE} Computer
  Society, 2018.

\bibitem{BCLQ18}
D.~Beaulaton, I.~Cristescu, A.~Legay, and J.~Quilbeuf.
\newblock A modeling language for security threats of iot systems.
\newblock In F.~Howar and J.~Barnat, editors, {\em Formal Methods for
  Industrial Critical Systems}, pages 258--268, Cham, 2018. Springer
  International Publishing.

\bibitem{bcmjk:07}
N.~Bena{\"i}ssa, D.~Cansell, and D.~M{\'e}ry.
\newblock Integration of security policy into system modeling.
\newblock In J.~Julliand and O.~Kouchnarenko, editors, {\em B 2007: Formal
  Specification and Development in B}, pages 232--247, Berlin, Heidelberg,
  2006. Springer Berlin Heidelberg.

\bibitem{cmt:12}
D.~M. Cappelli, A.~P. Moore, and R.~F. Trzeciak.
\newblock {\em {The CERT Guide to Insider Threats: How to Prevent, Detect, and
  Respond to Information Technology Crimes (Theft, Sabotage, Fraud)}}.
\newblock SEI Series in Software Engineering. Addison-Wesley Professional, 1
  edition, Feb. 2012.

\bibitem{suc:16}
CHIST-ERA.
\newblock Success: Secure accessibility for the internet of things, 2016.
\newblock http://www.chistera.eu/projects/success.

\bibitem{DBLP:conf/cav/EsparzaLNNSS13}
J.~Esparza, P.~Lammich, R.~Neumann, T.~Nipkow, A.~Schimpf, and J.~Smaus.
\newblock A fully verified executable {LTL} model checker.
\newblock In N.~Sharygina and H.~Veith, editors, {\em Computer Aided
  Verification - 25th International Conference, {CAV} 2013, Saint Petersburg,
  Russia, July 13-19, 2013. Proceedings}, volume 8044 of {\em Lecture Notes in
  Computer Science}, pages 463--478. Springer, 2013.

\bibitem{gdmgampd:18}
G.~Gonzalez-Granadillo, S.~Dubus, A.~Motzek, J.~Garcia-Alfaro, E.~Alvarez,
  M.~Merialdo, S.~Papillon, and H.~Debar.
\newblock Dynamic risk management response system to handle cyber threats.
\newblock {\em Future Generation Computer Systems}, 83:535--552, 2018.

\bibitem{DBLP:conf/itp/HaftmannKKN13}
F.~Haftmann, A.~Krauss, O.~Kuncar, and T.~Nipkow.
\newblock Data refinement in isabelle/hol.
\newblock In S.~Blazy, C.~Paulin{-}Mohring, and D.~Pichardie, editors, {\em
  Interactive Theorem Proving - 4th International Conference, {ITP} 2013,
  Rennes, France, July 22-26, 2013. Proceedings}, volume 7998 of {\em Lecture
  Notes in Computer Science}, pages 100--115. Springer, 2013.

\bibitem{hhs:86}
J.~He, C.~A.~R. Hoare, and J.~W. Sanders.
\newblock Data refinement refined.
\newblock In B.~Robinet and R.~Wilhelm, editors, {\em ESOP}, volume 213 of {\em
  Lecture Notes in Computer Science}, pages 187--196. Springer, 1986.

\bibitem{jac:89}
J.~Jacob.
\newblock On the derivation of secure components.
\newblock In {\em IEEE Security and Privacy}, pages 242--247. IEEE, 1989.

\bibitem{kam:16b}
F.~Kamm\"uller.
\newblock Isabelle modelchecking for insider threats.
\newblock In {\em Data Privacy Management, DPM’16, 11th Int. Workshop},
  volume 9963 of {\em LNCS}. Springer, 2016.
\newblock Co-located with ESORICS’16.

\bibitem{kam:18b}
F.~Kamm\"uller.
\newblock Attack trees in isabelle.
\newblock In {\em 20th International Conference on Information and
  Communications Security, ICICS2018}, volume 11149 of {\em LNCS}. Springer,
  2018.

\bibitem{kam:18a}
F.~Kamm\"uller.
\newblock Formal modeling and analysis of data protection for gdpr compliance
  of iot healthcare systems.
\newblock In {\em IEEE Systems, Man and Cybernetics, SMC2018}. IEEE, 2018.

\bibitem{kam:18smc}
F.~Kamm\"uller.
\newblock Isabelle infrastructure framework with iot healthcare s\&p
  application, 2018.
\newblock Available at \url{https://github.com/flokam/IsabelleAT}.

\bibitem{kam:19a}
F.~Kamm\"uller.
\newblock Combining secure system design with risk assessment for iot
  healthcare systems.
\newblock In {\em Workshop on Security, Privacy, and Trust in the IoT,
  SPTIoT’19, colocated with IEEE PerCom}. IEEE, 2019.

\bibitem{kam:19git}
F.~Kamm\"uller.
\newblock Isabelle infrastructure framework and rr-cycle with iot healthcare
  s\&p application, 2019.
\newblock Available at \url{https://github.com/flokam/IsabelleAT}.

\bibitem{kk:16}
F.~Kamm\"uller and M.~Kerber.
\newblock Investigating airplane safety and security against insider threats
  using logical modeling.
\newblock In {\em IEEE Security and Privacy Workshops, Workshop on Research in
  Insider Threats, WRIT'16}. IEEE, 2016.

\bibitem{kkp:16}
F.~Kamm\"uller, M.~Kerber, and C.~Probst.
\newblock Towards formal analysis of insider threats for auctions.
\newblock In {\em 8th ACM CCS International Workshop on Managing Insider
  Security Threats, MIST’16}. ACM, 2016.

\bibitem{knp:16}
F.~Kamm\"uller, J.~R.~C. Nurse, and C.~W. Probst.
\newblock Attack tree analysis for insider threats on the {IoT} using
  {I}sabelle.
\newblock In {\em Human Aspects of Information Security, Privacy, and Trust -
  Fourth International Conference, {HAS} 2015, Held as Part of {HCI}
  International 2016, Toronto}, Lecture Notes in Computer Science. Springer,
  2016.
\newblock Invited paper.

\bibitem{kop:18}
F.~Kamm\"uller, O.~O. Ogunyanwo, and C.~W. Probst.
\newblock Using fusion/uml for iot architecures for healthcare applications.
\newblock {\em arXiv}, https://arxiv.org/abs/1901.02426, 2018.

\bibitem{kp:14}
F.~Kamm\"uller and C.~W. Probst.
\newblock Combining generated data models with formal invalidation for insider
  threat analysis.
\newblock In {\em IEEE Security and Privacy Workshops (SPW)}. IEEE, 2014.

\bibitem{kp:16}
F.~Kamm\"uller and C.~W. Probst.
\newblock Modeling and verification of insider threats using logical analysis.
\newblock {\em IEEE Systems Journal, Special issue on Insider Threats to
  Information Security, Digital Espionage, and Counter Intelligence},
  11(2):534--545, 2017.

\bibitem{klpbb:14}
S.~Kanav, P.~Lammich, and A.~Popescu.
\newblock A conference management system with verified document
  confidentiality.
\newblock In A.~Biere and R.~Bloem, editors, {\em Computer Aided Verification},
  pages 167--183, Cham, 2014. Springer International Publishing.

\bibitem{DBLP:journals/jacm/Kobayashi13}
N.~Kobayashi.
\newblock Model checking higher-order programs.
\newblock {\em J. {ACM}}, 60(3):20:1--20:62, 2013.

\bibitem{kmswp:16}
A.~Kosba, A.~Miller, E.~Shi, Z.~Wen, and C.~Papamanthou.
\newblock Hawk: The blockchain model of cryptography and privacy-preserving
  smart contracts.
\newblock In {\em IEEE Symposium on Security and Privacy}, pages 839--858.
  IEEE, 2016.

\bibitem{DBLP:conf/fase/0012SRYHBRS18}
R.~Kumar, S.~Schivo, E.~Ruijters, B.~M. Yildiz, D.~Huistra, J.~Brandt,
  A.~Rensink, and M.~Stoelinga.
\newblock Effective analysis of attack trees: {A} model-driven approach.
\newblock In A.~Russo and A.~Sch{\"{u}}rr, editors, {\em Fundamental Approaches
  to Software Engineering, 21st International Conference, {FASE} 2018, Held as
  Part of the European Joint Conferences on Theory and Practice of Software,
  {ETAPS} 2018, Thessaloniki, Greece, April 14-20, 2018, Proceedings.}, volume
  10802 of {\em Lecture Notes in Computer Science}, pages 56--73. Springer,
  2018.

\bibitem{mar:19}
D.~Marmsoler.
\newblock {\em Towards Verified Blockchain Architectures: A Case Study on
  Interactive Architecture Verification}, pages 204--223.
\newblock 05 2019.

\bibitem{mcl:94}
J.~McLean.
\newblock A general theory of composition for trace sets closed under selective
  interleaving functions.
\newblock In {\em In Proc. IEEE Symposium on Security and Privacy}, pages
  79--93, 1994.

\bibitem{mor:09}
C.~Morgan.
\newblock The shadow knows: Refinement and security in sequential programs.
\newblock {\em Sci. Comput. Program.}, 74(8):629--653, 2009.

\bibitem{ml:98}
A.~C. Myers and B.~Liskov.
\newblock Complete, safe information flow with decentralized labels.
\newblock In {\em Proceedings of the IEEE Symposium on Security and Privacy}.
  IEEE, 1999.

\bibitem{DBLP:conf/cpp/0001HN13}
A.~Popescu, J.~H{\"{o}}lzl, and T.~Nipkow.
\newblock Formalizing probabilistic noninterference.
\newblock In G.~Gonthier and M.~Norrish, editors, {\em Certified Programs and
  Proofs - Third International Conference, {CPP} 2013, Melbourne, VIC,
  Australia, December 11-13, 2013, Proceedings}, volume 8307 of {\em Lecture
  Notes in Computer Science}, pages 259--275. Springer, 2013.

\bibitem{Schneier.102}
B.~Schneier.
\newblock {\em Secrets and Lies: Digital Security in a Networked World}.
\newblock John Wiley \& Sons, 2004.

\bibitem{sp:07}
N.~Stouls and M.-L. Potet.
\newblock Security policy enforcement through refinement process.
\newblock In J.~Julliand and O.~Kouchnarenko, editors, {\em B 2007: Formal
  Specification and Development in B}, pages 216--231, Berlin, Heidelberg,
  2006. Springer Berlin Heidelberg.

\bibitem{vnn:14}
R.~Vigo, F.~Nielsen, and H.~R. Nielsen.
\newblock Automated generation of attack trees.
\newblock In {\em 27th Computer Security Foundations Symposium, CSF'14}. IEEE,
  2014.

\end{thebibliography}
%%% -*-BibTeX-*-
%%% Do NOT edit. File created by BibTeX with style
%%% ACM-Reference-Format-Journals [18-Jan-2012].

%% Appendix
\appendix
%\section{Appendix}
%
%Text of appendix \ldots
\section{Background}
\label{sec:back}
This section provides an overview of the current extension of the Isabelle Infrastructure
framework in relation to previous works and how it integrates the Refinement-Risk cycle (Section
\ref{sec:isa}). It also summarizes the formalization of the existing theories for Kripke structures 
and the temporal logic CTL (Section \ref{sec:kripke}), as well as the attack tree formalisation
and Correctness and Completeness theorems (Section \ref{sec:at}). 
Finally, Section \ref{sec:hcapp} presents the IoT Healthcare system -- the case study on which 
the Refinement-Risk cycle is validated in this paper.

\subsection{Isabelle Infrastructure Framework}
\label{sec:isa}
Isabelle is a generic Higher Order Logic (HOL) proof assistant. Its generic
aspect allows the embedding of so-called object-logics as new theories
on top of HOL. There are sophisticated proof tactics available to support 
reasoning: simplification, first-order resolution, and special macros to support
arithmetic amongst others.
Object-logics are added to Isabelle using constant and type definitions
forming a so-called {\it conservative extension}. That is, no 
inconsistency can be introduced: new types are defined as 
subsets of existing types; properties are proved using a one-to-one 
relationship to the new type from properties of the existing type.
The use of HOL has the advantage that it enables expressing
even the most complex application scenarios, conditions, and logical
requirements. Isabelle enables the analysis of meta-theory, that is,
we can prove theorems {\it in} an object logic but also {\it about} it. 

This allows the building of telescope-like structures in which a meta-theory
at a lower level embeds a more concrete ``application'' at a higher level.
Properties are proved at each level.
Interactive proof is used to prove these 
properties but the meta-theory can be applied to immediately produce results.
Figure \ref{fig:theorystruc} in Section \ref{sec:intro}
gives an overview of the Isabelle Infrastructure 
framework with its layers of object-logics -- each level below embeds the one
above.
%\begin{figure}[h!]
%\begin{center}
%\includegraphics[scale=.4]{theory_structure}
%\end{center}
%%\vspace{-.5cm}
%\caption{Generic framework for infrastructures with model transformations.}
%\label{fig:theorystruc}
%\end{figure}

The Isabelle Infrastructure framework has been created initially for the modeling 
and analysis of Insider threats \cite{kp:16}. Its use has been validated on the 
most well-known insider threat patterns identified by the CERT-Guide to Insider 
threats \cite{cmt:12}. 
More recently, this Isabelle framework has been successfully applied to realistic
case studies of insider attacks in airplane safety \cite{kk:16} and on auction protocols
\cite{kkp:16}. These larger case studies as well as complementary work on the analysis
of Insider attacks on IoT infrastructures, e.g. \cite{knp:16},
have motivated the extension of the original framework by Kripke structures and 
temporal logic \cite{kam:16b} 
as well as a formalisation of attack trees \cite{kam:18b}.
Recently, GDPR compliance verification has been demonstrated \cite{kam:18a}.

\subsection{Kripke Structures and CTL}
\label{sec:kripke}
Kripke structures and CTL model state based systems and enable analysis
of properties under dynamic state changes.
A state transition relates snapshots of systems which are the states. 
The temporal logic CTL then enables expressing security and privacy
properties. 

In Isabelle, the system states and their transition relation are defined as a 
class called \texttt{state} containing an abstract constant \texttt{state\_transition}. 
It introduces the syntactic infix notation \texttt{I \ttrelI\, I'} to denote 
that system state \texttt{I} and \texttt{I'} are in this relation over an arbitrary 
(polymorphic) type $\ttsigma$. The operator \texttt{::} is a type judgement to 
coerce the type variable $\ttsigma$ into the class type. The arrow \texttt{\ttfun}
is the operator for functions on types and \texttt{bool} is the HOL inbuilt type
of truth values \texttt{true} and \texttt{false}.
\begin{ttbox}
 {\bf{class}} state =    
 {\bf{fixes}} state_transition :: (\ttsigma :: type) \ttfun \ttsigma \ttfun bool       
 ("_  \ttrelI _")
\end{ttbox}
The above class definition lifts Kripke structures and CTL to 
a general level.
The definition of the inductive relation is given by a set of specific rules
which are, however, %not necessary to define the notion of a Kripke structure
%and attack trees. They are 
part of an application like infrastructures (Section \ref{sec:infra}).
Branching time temporal logic CTL %has been integrated as part of the
%Isabelle Insider framework \cite{kam:16b} built 
is defined in general over Kripke structures with arbitrary state transitions 
and can later be applied to suitable theories, like infrastructures. 
%and state transitions and applied to the Insider theory. We generalise
%this theory here using the above class definition and thereby lifting
%Kripke structures and CTL to a generic level.

Based on the generic state transition $\ttrelI$ of the type class \texttt{state},
the CTL-operators \texttt{\sf EX} and \texttt{\sf AX} express that property $f$ 
holds in some or all next states, respectively.
%\begin{ttbox}
%{\sf AX} \ttf \ttequiv \{ s. \{f0. s \ttrelI f0 \} \ttsubseteq \ttf \}
%{\sf EX} \ttf \ttequiv \{ s. \ttexists f0 \ttin \ttf. s \ttrelI f0 \}
%\end{ttbox}
The CTL formula \texttt{\sf AG} $f$ means that on all paths branching from 
a state $s$ the formula $f$ is always true (\texttt{\sf G} stands for `globally'). 
It can be defined using the Tarski fixpoint theory by applying the greatest 
fixpoint operator.
%\begin{ttbox}
% {\sf AG} \ttf \ttequiv gfp(\ttlam Z. \ttf \ttcap {\sf AX} Z)
%\end{ttbox}
In a similar way, the other CTL operators are defined. 
The formal Isabelle definition of what it means that formula 
$f$ holds in a Kripke structure \texttt{M} can be
stated as: the initial states of the Kripke structure \texttt{init M} 
need to be contained in the set of all states \texttt{states M} 
that imply $f$.
\begin{ttbox}
 M \ttvdash f \ttequiv  init M \ttsubseteq \{ s \ttin states M. s \ttin f \}
\end{ttbox}
In an application, the set of states of the Kripke structure is defined 
as the set of states reachable by the infrastructure state transition from 
some initial state, say \texttt{ex\_scenario}.
\begin{ttbox}
  ex_states \ttequiv \{ I. ex_scenario \ttrelIstar  I \}
\end{ttbox}
The relation \texttt{\ttrelIstar} is the reflexive transitive closure -- an operator
supplied by the Isabelle theory library -- applied to the relation \texttt{\ttrelI}.

The \texttt{Kripke} constructor combines the constituents initial state and state set.
% and state transition relation \texttt{\ttrelI}. %and labelling function \texttt{L}.
\begin{ttbox}
 ex_Kripke \ttequiv Kripke ex_states \{ex_scenario\} 
\end{ttbox}
%Properties in HOL are given as predicates. Alternatively, they can
%be seen a sets in a view which is often called {\it predicate transformer}
%semantics. 
%In HOL, sets are defined as predicates -- so these concepts coincide.
%Given some \texttt{property} -- a predicate over states (or equally a set of 
%states) --  we can thus then for example try to 
%prove that there is a path ({\sf E}) to a state in which the property 
%eventually holds (in the {\sf F}uture) by attempting the following proof 
%in Isabelle.
In Isabelle, the concept of sets and predicates coincide (more precisely they are isomorphic)
\footnote{In general, this is often referred to as {\it predicate transformer 
semantics.}}.
Thus a \texttt{property} is a predicate over states which is equal to a set of 
states. For example, we can then try to prove that there is a path ({\sf E}) 
to a state in which the property eventually holds (in the {\sf F}uture) by 
starting the following proof in Isabelle.
\begin{ttbox}
 ex_Kripke \ttvdash {\sf EF} property 
\end{ttbox}
Since \texttt{property} is a set of states, and the temporal operators
are predicate transformers, that is, transform sets of states to sets of states, 
the resulting {\sf EF} \texttt{property} is also a set of states -- and hence 
again a property.

\subsection{Attack Trees in Isabelle}
\label{sec:at}
\begin{figure}[h!]
\begin{center}
\includegraphics[scale=.25]{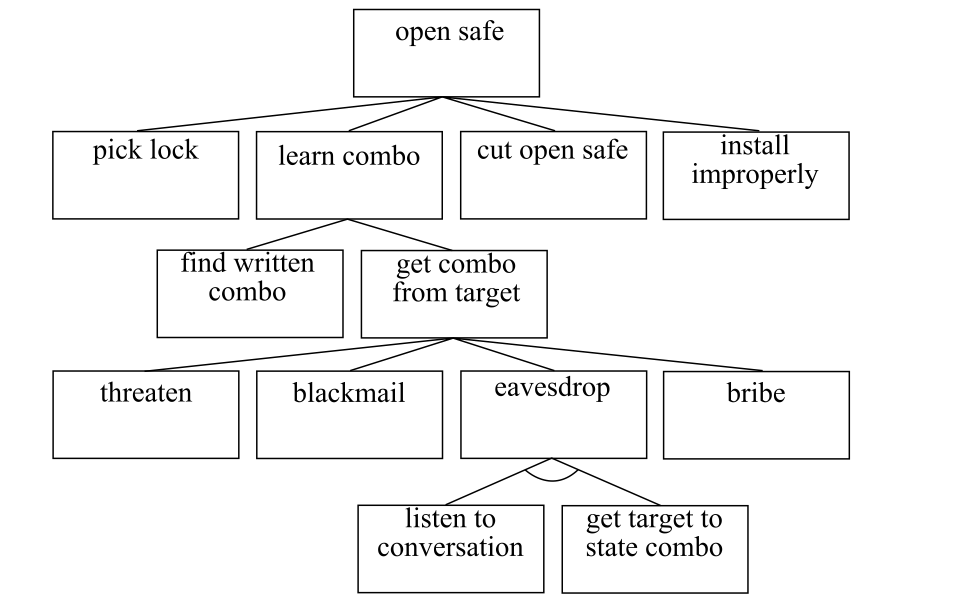}
\end{center}
\caption{Attack tree example illustrating disjunctive nodes for alternative 
attacks refining the attack ``open safe''. Near the leaves there is also a 
conjunctive node ``eavesdrop''.}
\label{fig:atex}
\end{figure}

Attack trees \cite{Schneier.102} are a graphical language for the analysis 
and quantification of attacks. If the root represents an attack, 
its children represent the sub-attacks. 
Leaf nodes are the basic attacks; other
nodes of attack trees represent sub-attacks.
%can be and-nodes and or-nodes that combine their 
%sub-trees either in a conjunctive or in a disjunctive manner. 
Sub-attacks can be alternatives for reaching the goal (disjunctive node) or 
they must all be completed to reach the goal (conjunctive node). 
%\TODO{Insert explanation to previous reviewer about parallev vs seq AND}
% Done see below at the point of the definition of the is_attack_tree predicate
Figure \ref{fig:atex} is an example of an attack tree taken from a textbook
\cite{Schneier.102} illustrating the attack of opening a safe.
Nodes can be adorned with attributes, for example costs of attacks
or probabilities which allows quantification of attacks (not used 
in the example).

The following datatype definition \texttt{attree} defines attack trees.
Isabelle allows recursive datatype definitions similar to the programming languages
Haskell or ML. A datatype is given by a ``\texttt{|}'' separated sequence of
possible cases each of which consists of a constructor name, the types of inputs to this
constructor, and optionally a pretty printing syntax definition.
The simplest case of an attack tree is a base attack.
The principal idea is that base attacks are defined by a pair of
state sets representing the initial states and the {\it attack property}
-- a set of states characterized by the fact that this property holds
for them. 
Attacks can also be combined as the conjunction or disjunction of other attacks. 
The operator $\oplus_\vee$ creates or-trees and $\oplus_\wedge$ creates and-trees.
And-attack trees $l \ttattand s$ and or-attack trees $l \ttattor s$ 
consist of a list of sub-attacks -- again attack trees. 
\begin{ttbox}
{\bf datatype} (\ttsigma :: state)attree = 
  BaseAttack (\ttsigma set)\tttimes(\ttsigma set) ("\ttcalN (_)") 
| AndAttack (\ttsigma attree)list (\ttsigma set)\tttimes(\ttsigma set) ("_ {\ttattand{(\_)}}")
| OrAttack  (\ttsigma attree)list (\ttsigma set)\tttimes(\ttsigma set) ("_ {\ttattor{(\_)}}")
\end{ttbox}
The attack goal 
is given by the pair of state sets on the right of the operator 
\texttt{\ttcalN}, $\oplus_\vee$ or $\oplus_\wedge$, respectively. A corresponding 
projection operator is defined as the function \texttt{attack}.

\begin{figure*}
\begin{center}
\includegraphics[scale=.3]{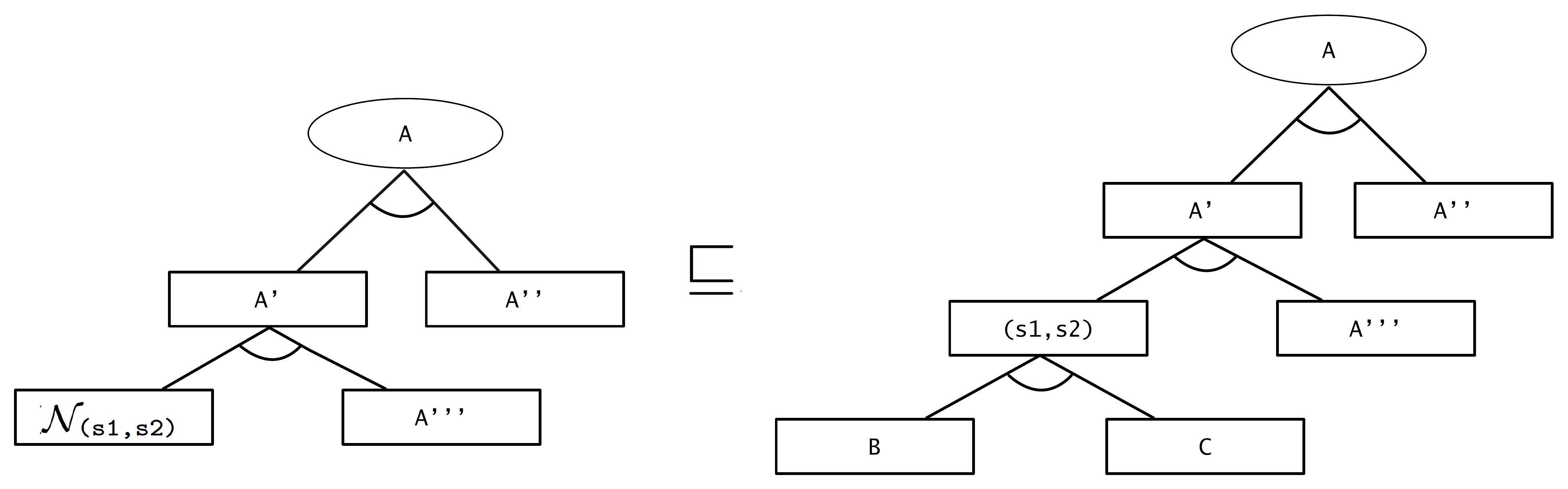}
\end{center}
%\vspace{-.5cm}
\caption{Attack tree example illustrating refinement of an and-subtree.}
\label{fig:ref}
\end{figure*}

When we develop an attack tree, we proceed from an abstract attack, given
by an attack goal, by breaking it down into a series of sub-attacks. This
proceeding corresponds to a process of {\it refinement}. 
The attack tree calculus \cite{kam:18b} provides a notion of attack tree refinement
elegantly expressed as the infix operator $\sqsubseteq$. 
Note that this refinement is different from the notion of system refinement that
will be presented later in this paper.
The intuition of developing an attack tree by refinement from the root to the leaves
is illustrated  in Figure \ref{fig:ref} (the formal definition is in \cite{kam:18b}).
The example attack tree on the left side has a leaf that is expanded by the refinement 
into an and-attack with two steps.
Refinement of attack trees defines the stepwise process of expanding abstract
attacks into more elaborate attacks only syntactically. There is no guarantee
that the refined attack is possible if the abstract one is, nor vice-versa.
The attack tree calculus \cite{kam:18b} formalizes the semantics of attack trees
on Kripke structures and CTL enabling rigorous judgement whether such
syntactic refinements represent possible attacks.

A valid attack, intuitively, is one which is fully refined into fine-grained
attacks that are feasible in a model. The general model provided is
a Kripke structure, i.e., a set of states and a generic state transition.
Thus, feasible steps in the model are single steps of the state transition.
They are called valid base attacks.
The composition of sequences of valid base attacks into and-attacks yields
again valid attacks if the base attacks line up with respect to the states
in the state transition. If there are different valid attacks for the same
attack goal starting from the same initial state set, these can be 
summarized in an or-attack. The formal definition \cite{kam:18b} is given in the table
in Figure \ref{fig:isatttree}.
\begin{figure*}
\begin{ttbox}
{\bf fun} is_attack_tree :: [(\ttsigma :: state) attree] \ttfun bool  ("\ttvdash_") 
{\bf where} 
  att_base:  \ttvdash \ttcalN{s} = \ttforall x \ttin fst s. \ttexists y \ttin snd s. x  \ttrelI y  
| att_and: \ttvdash (As :: (\ttsigma::state attree list)) \ttattand{s} = 
           case As of
             [] \ttfun (fst s \ttsubseteq snd s)
           |  [a] \ttfun \ttvdash a \ttand attack a = s 
           |  a \# l \ttfun \ttvdash a \ttand fst(attack a) = fst s 
                         \ttand \ttvdash l \ttattand{\texttt{(snd(attack a),snd(s))}} 
| att_or: \ttvdash (As :: (\ttsigma::state attree list)) \ttattor{s} = 
          case As of 
             [] \ttfun (fst s \ttsubseteq snd s) 
          | [a] \ttfun \ttvdash a \ttand fst(attack a) \ttsupseteq fst s \ttand snd(attack a) \ttsubseteq snd s
          | a \# l \ttfun \ttvdash a \ttand fst(attack a) \ttsubseteq fst s \ttand snd(attack a) \ttsubseteq snd s
                       \ttand \ttvdash l \ttattor{\texttt{(fst s - fst(attack a),snd s)}}
\end{ttbox}
\caption{Definition of attack tree validity as one recursive predicate.}\label{fig:isatttree}
\end{figure*}
The semantics of attack trees is described by this one recursive function. Since the definition
can be given as a recursive function, Isabelle code generation is applicable: an executable
decision procedure for attack tree validity can be automatically generated in various programming 
languages, for example, Scala.

Adequacy of the semantics is proved in \cite{kam:18b} by proving correctness and completeness.
The following correctness theorem shows that
if \texttt{A} is a valid attack on property \texttt{s} starting from
initial states described by \texttt{I}, then from all states in \texttt{I} 
there is a path to the set of states fulfilling \texttt{s} in the 
corresponding Kripke structure.
\begin{ttbox}
{\bf theorem} AT_EF: \ttvdash A :: (\ttsigma :: state) attree) \ttImp 
 (I, s) = attack A \ttImp 
 Kripke \{t . \ttexists i \ttin I. i \ttrelI^* t\} I \ttvdash {\sf EF} s
\end{ttbox}
The inverse direction of theorem \texttt{AT\_EF} is a completeness
theorem: if states described by predicate \texttt{s} can be reached from a 
finite nonempty set of initial states \texttt{I} in 
a Kripke structure, then there exists a valid attack tree for the attack 
\texttt{(I,s)}.
\begin{ttbox} 
{\bf theorem} Completeness: I \ttneq \{\} \ttImp finite I \ttImp
 Kripke \{t . \ttexists i \ttin I. i \ttrelI^* t\} I \ttvdash {\sf EF} s \ttImp
 \ttexists A :: (\ttsigma::state)attree. \ttvdash A \ttand (I, s) = attack A 
\end{ttbox}
Correctness and Completeness are proved in Isabelle \cite{kam:18b, kam:18smc}.
They are not just necessary proofs on the attack tree semantics but the theorems
allow easy transformation of properties between the embedded notions
of attack tree validity $\vdash$ and CTL formulas like {\sf EF}.
The relationship between these notions can be applied to case studies. That is, if we 
apply attack tree refinement to spell out an abstract attack tree for attack \texttt{s} 
into a valid attack sequence, we can apply theorem \texttt{AT\_EF} and can immediately 
infer that {\sf EF} \texttt{s} holds. Vice versa, the theorem Completeness can be applied
to directly infer the existence of an attack tree from the former.

\subsection{Edge Computing: IoT Healthcare System}
\label{sec:hcapp}
\begin{figure}[h]
\begin{center}
\includegraphics[scale=.19]{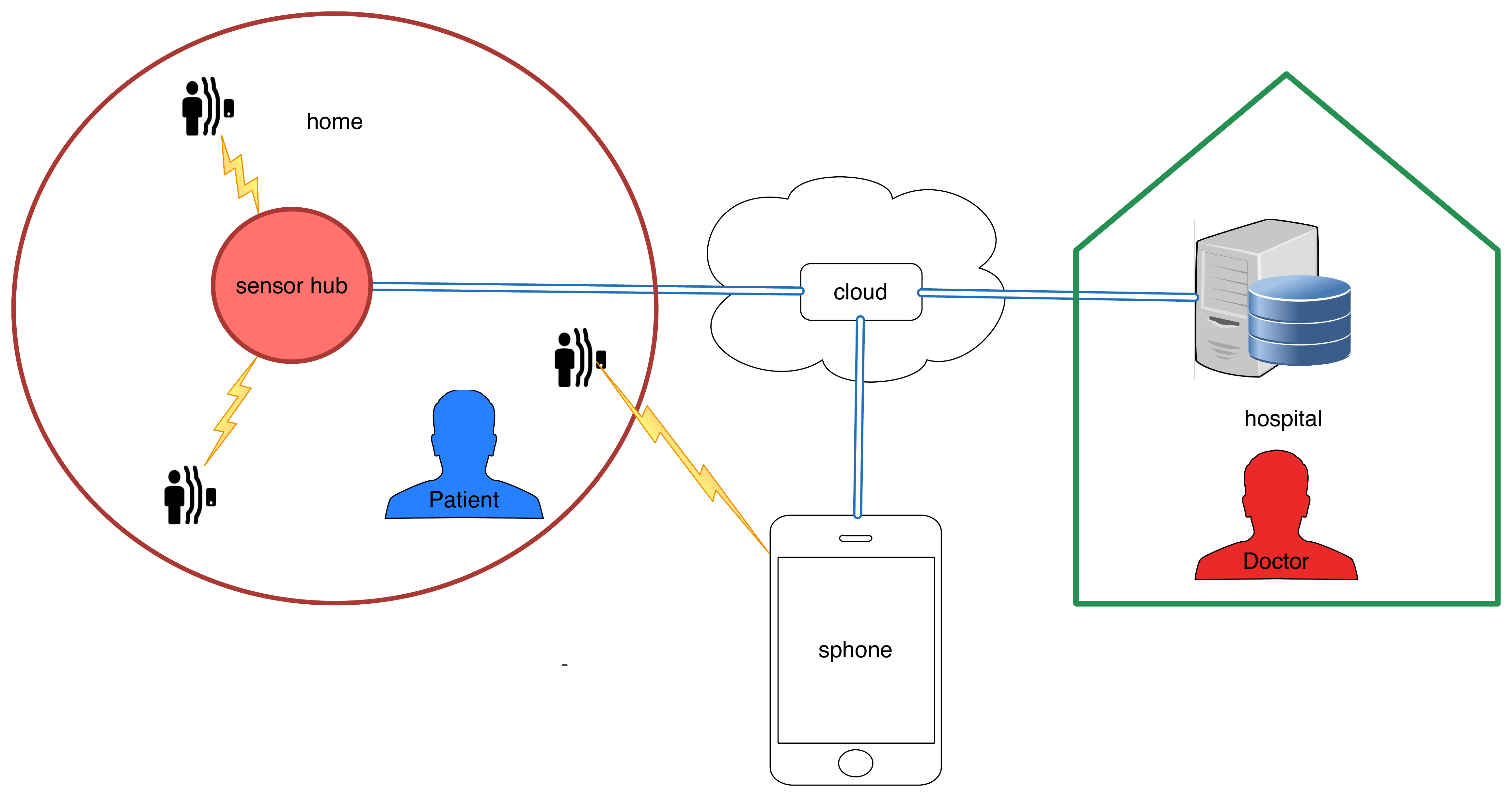}
\caption{IoT healthcare monitoring system for SUCCESS project \cite{suc:16}}\label{fig:iot}
\end{center}
%\vspace{-.5cm}
\end{figure}

% Describe the model of the IoT Healthcare system
%\subsection{SUCCESS Security and Privacy for IoT Healtcare}
%\label{sec:success}
% short summary of project and goals - could be part of some backgrond section
% or part of the intro to case study
%First, we introduce the application domain of transparent security and privacy of
The example of an IoT healthcare systems is from the CHIST-ERA project SUCCESS \cite{suc:16}
on monitoring Alzheimer's patients. % (Section \ref{sec:success}). 
Figure \ref{fig:iot} illustrates the system architecture where data collected by sensors 
in the home or via a smartphone helps monitoring bio markers of the patient. The data 
collection is in a cloud based server to enable hospitals (or scientific institutions) 
to access the data which is controlled via the smartphone.
It is a typical edge network application: the smartphone and the sensor hub in the 
home are typical edge devices that are capable of doing processing data without uploading
to the cloud server.

\end{document}